\documentclass[twocolumn]{pasj02} 
\usepackage[switch,mathlines]{lineno} 
\usepackage{color}
\usepackage{rotating}

\newcommand{\fukushima}{\textup{Fukushima et al. in preparation}}
\newcommand{\atomdb}{\textup{AtomDB}}

\jyear{2025}
\Received{}
\Accepted{}

\begin{document} 

\title{X-ray line diagnostics of the multi-phase gas in the Centaurus cluster core with XRISM/Resolve}

\author{
Marie~\textsc{Kondo},\altaffilmark{1}\altemailmark\orcid{0009-0005-5685-1562} \email{m.kondo.366@ms.saitama-u.ac.jp} 
Kotaro~\textsc{Fukushima},\altaffilmark{2}\orcid{0000-0001-8055-7113} 
Kazunori~\textsc{Suda},\altaffilmark{3}
Anwesh~\textsc{Majumder},\altaffilmark{4,5}\orcid{0000-0002-3525-7186}
Kosuke~\textsc{Sato},\altaffilmark{6,1}\altemailmark\orcid{0000-0001-5774-1633} \email{ksksato@cc.kyoto-su.ac.jp} 
Kyoko~\textsc{Matsushita},\altaffilmark{3}\orcid{0000-0003-2907-0902}
Fran\c{c}ois~\textsc{Mernier},\altaffilmark{7,8,9,10}\orcid{0000-0002-7031-4772} 
Kazuhiro~\textsc{Nakazawa},\altaffilmark{11}\orcid{0000-0003-2930-350X}
Aurora~\textsc{Simionescu},\altaffilmark{5}\orcid{0000-0002-9714-3862}
Jean-Paul~\textsc{Breuer},\altaffilmark{12}\orcid{0000-0001-6131-4802}
Yasushi~\textsc{Fukazawa},\altaffilmark{12}\orcid{0000-0002-0921-8837}
Ryuichi~\textsc{Fujimoto},\altaffilmark{2}\orcid{0000-0002-2374-7073}
Isamu~\textsc{Hatsukade},\altaffilmark{13}\orcid{0000-0003-3518-3049}
Kokoro~\textsc{Hosogi},\altaffilmark{14}
Michael~\textsc{Loewenstein},\altaffilmark{8,9,10}\orcid{0000-0002-1661-4029}
Tom\'a\v{s}~\textsc{Pl\v{s}ek},\altaffilmark{15}\orcid{0000-0001-6411-3651}
Ming~\textsc{Sun},\altaffilmark{14}\orcid{0000-0001-5880-0703}
Misaki~\textsc{Urata},\altaffilmark{12}\orcid{0009-0008-3206-235X}
Norbert~\textsc{Werner},\altaffilmark{15}\orcid{0000-0003-0392-0120}
Noriko~Y.~\textsc{Yamasaki},\altaffilmark{2,16}\orcid{0000-0003-4885-5537}
and Yutaka~\textsc{Fujita}\altaffilmark{17}\orcid{0000-0003-0058-9719}
}

\altaffiltext{1}{Department of Physics, Saitama University, 255 Shimo-Okubo, Sakura-ku, Saitama, Saitama 338-8570, Japan}
\altaffiltext{2}{Institute of Space and Astronautical Science, JAXA, 3-1-1 Yoshinodai, Chuo-ku, Sagamihara, Kanagawa 252-5210, Japan}
\altaffiltext{3}{Department of Physics, Tokyo University of Science, 1-3 Kagurazaka, Shinjuku-ku, Tokyo 162-8601, Japan}
\altaffiltext{4}{Waterloo Centre for Astrophysics, University of Waterloo, 200 University Avenue West, Waterloo, Ontario N2L 3G1, Canada}
\altaffiltext{5}{SRON Space Research Organisation Netherlands, Niels Bohrweg 4, 2333 CA Leiden, The Netherlands}
\altaffiltext{6}{Department of Astrophysics and Atmospheric Sciences, Kyoto Sangyo University, Motoyama, Kamigamo, Kita-ku, Kyoto, Kyoto 603-8555, Japan}
\altaffiltext{7}{IRAP, CNRS, Université de Toulouse, CNES, UT3-UPS, Toulouse, France}
\altaffiltext{8}{Department of Astronomy, University of Maryland, College Park, Maryland 20742, USA}
\altaffiltext{9}{Goddard Space Flight Center, NASA, Greenbelt, Maryland 20771, USA}
\altaffiltext{10}{Center for Research and Exploration in Space Science and Technology, NASA/GSFC (CRESST II), Greenbelt, Maryland 20771, USA}
\altaffiltext{11}{Department of Physics, Nagoya University, Furo-cho, Chikusa-ku, Nagoya, Aichi 464-8602, Japan}
\altaffiltext{12}{Department of Physics, Hiroshima University, 1-3-1 Kagamiyama, Higashi-Hiroshima, Hiroshima 739-8526, Japan}
\altaffiltext{13}{Faculty of Engineering, University of Miyazaki, 1-1 Gakuen-Kibanadai-Nishi, Miyazaki, Miyazaki 889-2192, Japan}
\altaffiltext{14}{Department of Physics and Astronomy, University of Alabama in Huntsville, Huntsville, Alabama, USA}
\altaffiltext{15}{Department of Theoretical Physics and Astrophysics, Masaryk University, Kotl\'a\v{r}sk\'a 2, Brno 611 37, Czech}
\altaffiltext{16}{Department of Physics, The University of Tokyo, 7-3-1 Hongo, Bunkyo-ku, Tokyo 113-0033, Japan}
\altaffiltext{17}{Department of Physics, Tokyo Metropolitan University, 1-1 Minami-Osawa, Hachioji, Tokyo 192-0397, Japan}




\maketitle

\begin{abstract}

We report the multi-temperature structure of the intracluster medium (ICM) in the Centaurus cluster core observed with XRISM/Resolve. Thanks to its high energy resolution, Resolve enables us to measure fine structures of highly ionized emission lines from Si to Fe and to directly determine the excitation temperature and the ionization temperature from the emission line ratio diagnostics.
The observed spectrum in the Centaurus core is well-represented by a double-temperature thermal plasma at collisional ionization equilibrium state rather than an isothermal one. The line ratio diagnostics also support this biphasic temperature structure. 
Particularly, the observed line ratios show a trend of increasing ionization temperature with atomic mass, while the ionization and excitation temperatures of Fe show nearly the same temperature. 
The resultant line ratios, which are well-represented by the two temperatures ICM, $\sim 1.6$ and $\sim 3$ keV, are also fairly consistent with the expected numbers when assuming the radial single-temperature ICM was projected in the cluster core along the line of sight.
Due to the limited low-energy sensitivity of the Resolve with the gate valve closed, we investigated the effect of the cool component using the XMM-Newton/RGS spectrum, but it ultimately did not affect our results.
The observed flux ratio between the Fe \emissiontype{XXV} He $\alpha$ resonance and forbidden lines shows an about 20\% reduction, suggesting the presence of resonant scattering.
\end{abstract}
\setcounter{tocdepth}{4} 


\section{Introduction}

\begin{figure}[t]
\begin{center}
\includegraphics[width=80mm]{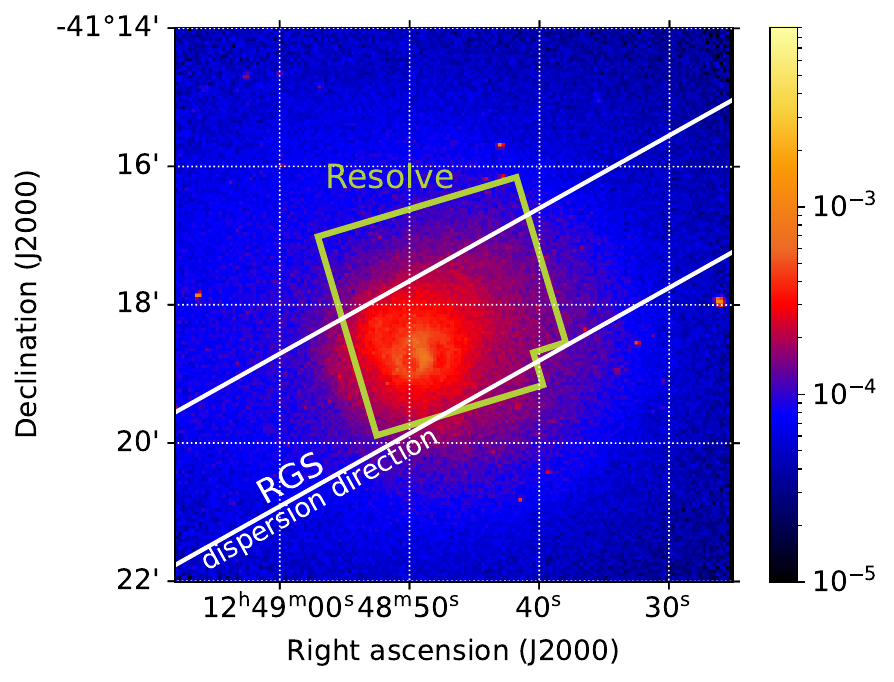}
\end{center}
\caption{The XRISM/Resolve FoV (green box) overlaid on the Chandra X-ray image in the 2.0--8.0 keV energy band for the Centaurus cluster core. The missing part of the FoV corresponds to the calibration pixel 12. The extraction region for the XMM-Newton/RGS spectral analysis is shown by white solid lines.
The data sets are available at https://doi.org/10.25574/cdc.500. {Alt text: One X-ray image.}
}
\label{fig:img_RGS} 
\end{figure}

Galaxy clusters are the largest gravitationally bound structures in the Universe. They consist of dark matter, hot gas, and galaxies, with the intracluster medium (ICM) being the dominant component that can be observed through electromagnetic radiation (e.g., \cite{1997ApJ...481..660I,1998ApJ...498..606S}). Their thermodynamic properties, including temperature structure and hydrostatic equilibrium, are important for estimating cluster masses, which directly constrain cosmological parameters (e.g., \cite{2001MNRAS.324..877A,2001PASJ...53..401M}).
Despite the fact that radiative cooling should be effective in the relatively dense environment of the cluster core region, X-ray observations have revealed that many galaxy clusters have cool cores, regions where the temperature stops decreasing and is maintained at a certain level. X-ray observations have indeed shown low-temperature gas in cool cores; however, the quantity of this gas is significantly lower than expected (e.g., \cite{2001A&A...365L.104P,2001A&A...365L..87T,2001A&A...365L..99K,2006PhR...427....1P}).
Active galactic nuclei (AGNs) are often found in the cores of galaxies, so they are often thought to be the heating sources (e.g., \cite{2007ARA&A..45..117M}).
The energy could be transferred to the surrounding ICM through sound waves (e.g., \cite{2006MNRAS.366..417F}), shocks (e.g., \cite{2015ApJ...805..112R}), cosmic rays (e.g., \cite{2013MNRAS.432.1434F}), and turbulence (e.g., \cite{2003ApJ...596L.139K}).
Understanding the temperature structure of cool-core clusters enables us to trace the history of cooling, heating, and gas mixing processes that shaped their cores. Recent X-ray CCD observations with Chandra, XMM-Newton, and Suzaku have revealed that multi-temperature models accurately represent the multi-phase ICM structure in cool cores (e.g., Abell~4059: \cite{2015A&A...575A..37M}, Hydra~A: \cite{2009A&A...493..409S}, Abell~496: \cite{2014A&A...570A.117G}). 

The temperature measurements of the ICM with X-ray CCD have mainly been derived from the continuum spectral shape, which is affected by uncertainties such as the effective area and background modeling. In addition, the coupling between the temperature of the continuum shape and abundances from emission lines makes it difficult to determine the accurate temperature due to insufficient energy resolution capability. In particular, the Fe-L complex around 1 keV has been reported to significantly affect the temperature measurements. The temperature measurements based on emission line ratios from different energy levels are less affected by such uncertainties. Before the Hitomi era, such temperature measurements were conducted only using strong emission lines. 

The high spectral resolution of Hitomi enabled precise diagnostics of the thermal and dynamical states of the ICM in the Perseus cluster.
\citet{2018PASJ...70...11H} performed the direct measurement of not only the electron temperature but also various ionization degrees of elements from Si to Fe using the Hitomi observation of the Perseus cluster. They also report the excitation temperature derived from the same ionization state, such as Ly$\beta$/Ly$\alpha$, and the ionization temperature from the different ionization state, such as Ly$\alpha$/He$\alpha$, from the emission line diagnostics.
While most line ratios are largely consistent with a single-temperature CIE plasma, small deviations are identified: the ionization temperature increases with atomic mass, and the excitation temperature is lower than the ionization temperature for Fe in the core.
These results agree with a multi-temperature structure projected along the line of sight using the temperature profile from Chandra.
The dynamical properties of the ICM were constrained by emission line widths of S, Ca, and Fe \citep{2018PASJ...70....9H}, 
measuring both the line-of-sight velocity dispersion of random motion, including turbulence, $\sigma_v$, and of the thermal motion of ions, $\sigma_\textup{th}$. The resultant ion temperature from the line width is consistent with the electron temperature within the 68\% confidence level.

Finally, the Fe \emissiontype{XXV} He$\alpha$ resonance line ($w$) was found to be suppressed toward the core, and its width was broader than the other Fe \emissiontype{XXV} triplet lines \citep{2018PASJ...70...10H}.
The expected turbulent velocities required to reproduce the observed fluxes of Fe He$\alpha$ $w$ (optically thick), He$\alpha$ forbidden ($z$), He$\beta$, and \emissiontype{XXVI} Ly$\alpha$ (optically thin) in Monte Carlo radiative-transfer simulations were consistent with direct measurements.

\begin{figure*}[t]
\begin{minipage}{1.0\hsize}
\begin{center}
\includegraphics[width=145mm]{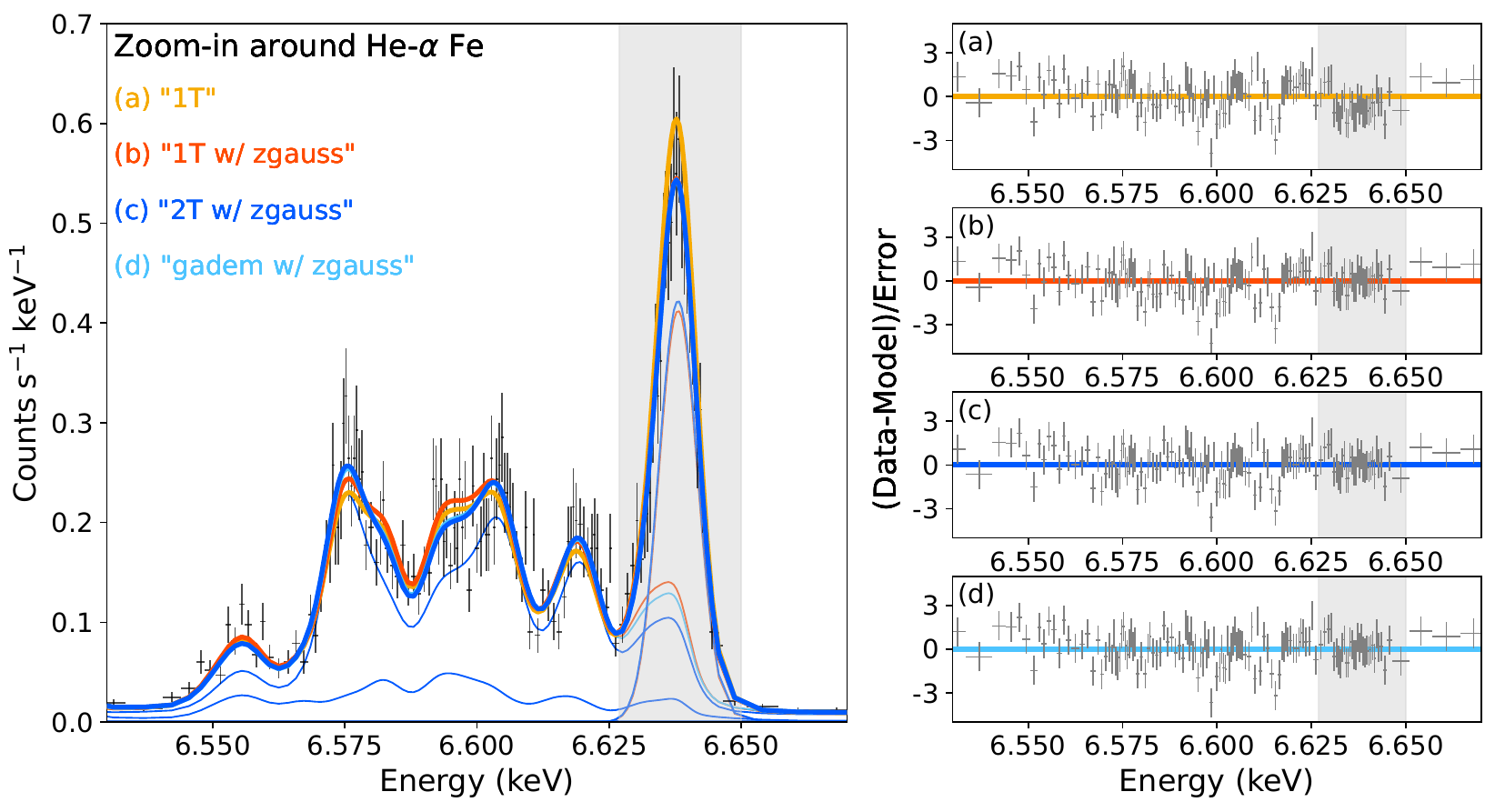}
\end{center}
\end{minipage}
\begin{minipage}{0.5\textwidth}
 \begin{center}
 \includegraphics[width=80mm]{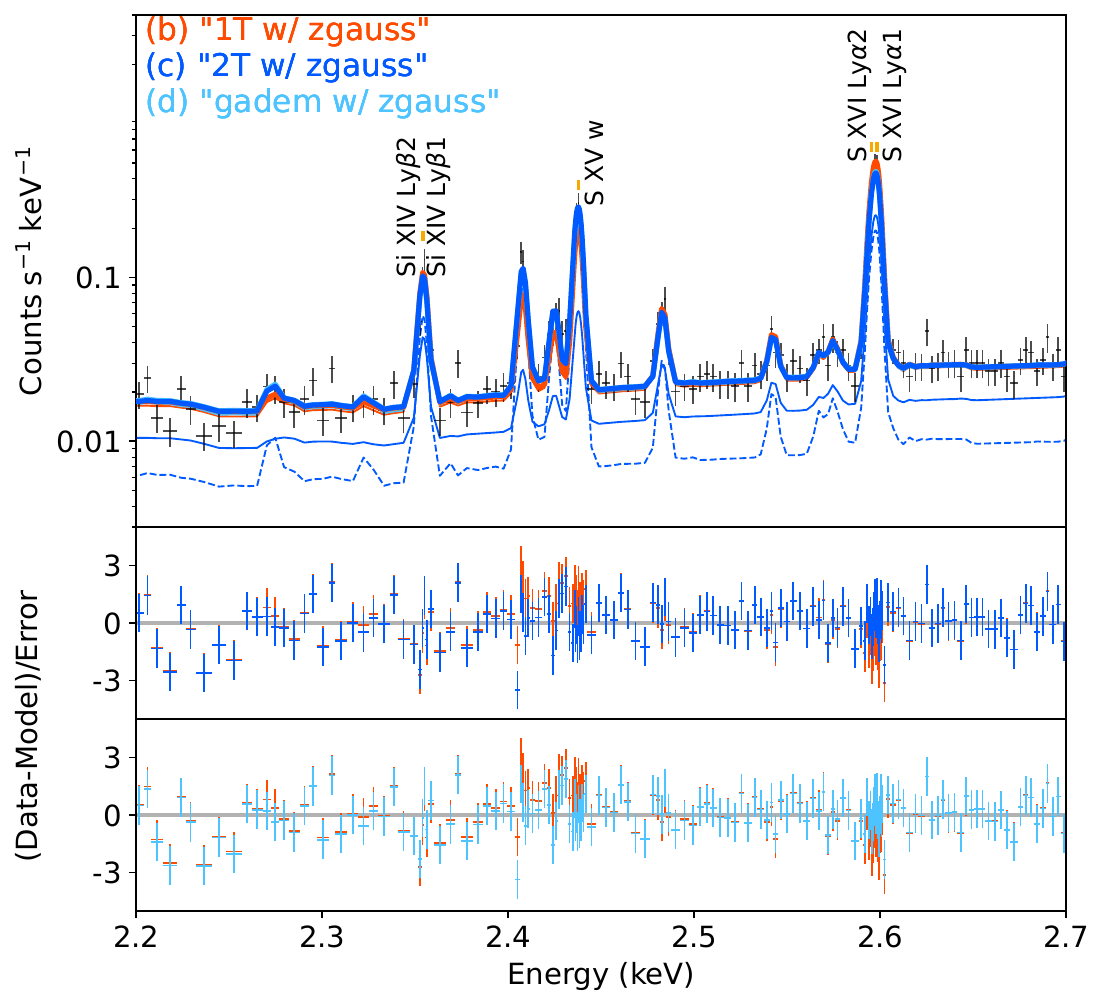}
 \end{center}
 \end{minipage}
 \begin{minipage}{0.5\textwidth}
\begin{center}
\includegraphics[width=80mm]{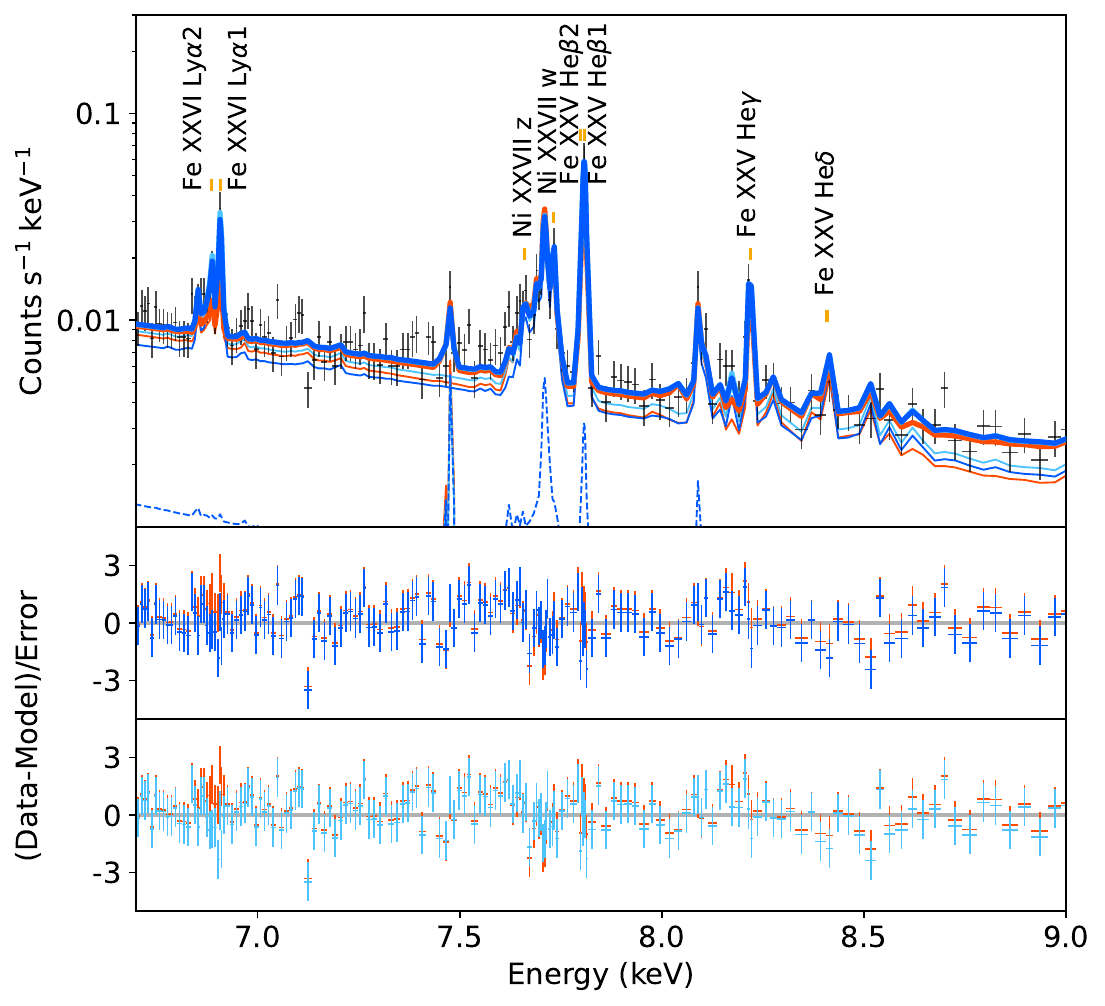}
\end{center}
\end{minipage}
\caption{Upper left panel shows the zoom-in spectra around Fe He$\alpha$ triplet for the Resolve entire region spectrum. The orange, red, blue, and light blue solid lines indicate the best-fit models: (a) ``1T'', (b) ``1T w/ zgauss'', (c) ``2T w/ zgauss'', and (d) ``gadem w/ zgauss'', respectively. The upper right panels show the residuals corresponding to models (a) to (d). The gray-shaded region shows the energy range around the Fe\emissiontype{XXV} resonance line. The lower panels show the zoom-in Resolve spectral fits in the 2.2--2.7 keV band (left) and the 6.7--9.0 keV band (right). The thin blue solid and dashed lines indicate the higher and lower temperature components of the ``2T w/ zgauss'' model, respectively. The middle panels show the residuals between the data and the ``1T w/ zgauss'' and ``2T w/ zgauss'' models. The bottom panels also show the residuals between the data and the ``1T w/ zgauss'' and ``gadem w/ zgauss'' models. {Alt text: Seven line graphs. The upper left graph shows X-ray spectra near the He $\alpha$ Fe line. The upper right plots display residuals for four test models. In the two plots shown below, the y-axis of the upper panel represents the count rate per kiloelectron volt, whereas the y-axes of the middle and lower panels show the residuals between the data and the model. }} 
\label{fig:spec_Resolve1T2T}
\end{figure*}

An X-ray microcalorimeter spectrometer, Resolve \citep{Kelley et al. JATIS 2025}, onboard XRISM satellite \citep{2025PASJ..tmp...28T} enables us to perform the high-resolution spectroscopy of the ICM in galaxy clusters. 
As for direct measurements of the ICM thermal structure, observations of Abell~2029 demonstrated an intriguing result.  \citet{2025PASJ...77S.254S} investigated the multi-temperature structure of Abell~2029 using line diagnostics. 
Detailed line diagnostics using the XRISM satellite have so far been carried out only for Abell~2029. In addition to the Perseus cluster observed with Hitomi and Abell~2029, it is essential to perform temperature measurements for clusters across other temperature ranges as well. For instance, intermediate-temperature clusters allow the use of not only Fe lines but also Si and S lines, which are predominantly emitted by cooler gas, for detailed line diagnostics.
The Centaurus cluster is one of the nearest X-ray bright clusters. The AGN of its central galaxy, NGC 4696, is quieter than that of the Perseus cluster. Its central temperature is lower than that of the Perseus cluster. In terms of AGN activity and ICM temperature, the Centaurus cluster is a suitable target for comparison with the Perseus cluster. Previous studies with Chandra have revealed complex dynamical structures in the central region \citep{2002MNRAS.331..273S}. These studies identified a plume-like structure extending approximately 60 arcseconds from the cluster center and found that the temperature increases sharply near the end of the plume, indicating the presence of a cold front.

The presence of multi-temperature plasma in the Centaurus cluster core has been reported \citep{1994PASJ...46L..55F,1999ApJ...525...58I}. 
Based on Chandra observations, \citet{2016MNRAS.457...82S} showed two-dimensional deviation maps of surface brightness, temperature, and metallicity relative to the azimuthally averaged distribution of the ICM. They found positive correlations between the low-temperature and high-surface-brightness regions, as well as between the metal-rich and high-surface-brightness regions. These spatial correlations were interpreted as evidence of gas motions induced by sloshing. 
\citet{2008MNRAS.385.1186S} found a cool component (0.3–0.45 keV) in the core of the Centaurus cluster based on the ratio of the Fe \emissiontype{XVII} 17.1 \AA\,  lines to the 15.0 \AA\, line and limits on O \emissiontype{VII} emission with the XMM-Newton RGS. They also reported that the amount of cool gas detected is significantly lower than that predicted by the cooling-flow model.
Other characteristics of the Centaurus cluster are the high Fe abundance ($\sim 1.5$ Solar, e.g., \cite{2002MNRAS.331..273S}). \citet{2022MNRAS.514.4222F} reported temperature and metal abundances using Chandra ACIS-S, XMM-Newton MOS, pn, and RGS data. In the innermost region within 5 arcsec from the cluster center, the RGS spectrum was well-represented by a three-temperature model, with the hottest component at $3.02\pm0.09$ keV and an emission-measure-weighted temperature of $1.6\pm0.8$ keV. 

XRISM observed Centaurus as one of the first targets during its Performance Verification (PV) phase, following the calibration phase. The first study presented its Resolve spectrum, as well as a map of its bulk and turbulent gas velocity measurements
\citep{2025Natur.638..365X}. This study uses the same deep dataset to examine the ICM temperature.  Details on abundances are studied in \citet{Mernier2025submitted} and \fukushima{}.
Throughout this paper, we performed the spectral analysis using the Xspec 12.14.1 package and used the \atomdb{} version 3.1.3 to reproduce CIE plasma spectra. We adopted the solar abundance table provided by \citet{2009LanB...4B..712L}. The Galactic hydrogen column density of $N_\textup{H} = 7.77 \times 10^{20}$\,\textup{cm}$^{-2}$ \citep{2016A&A...594A.116H} in the direction of the Centaurus cluster, the same as in \citet{2025Natur.638..365X}, was adopted for the spectral analysis, unless noted otherwise. Errors are given at the 1$\sigma$ (68$\%$) confidence level.

\begin{table*}
  \tbl{Summary of the resultant best-fit parameters with each spectral model.}{%
  \begin{tabular}{lccccccccc}
    \hline
 Model  && $kT_{\textup{mid}}$  & $kT_{\textup{low}}$  && $\textup{Norm}_{\textup{mid}}/\textup{Norm}_{\textup{low}}$ & $\chi^{2}/$d.o.f. \\
   && (keV) & (keV) && &\\    \hline 
        RGS            &             &           &             &            \\
   2T    & &  $1.85_{-0.03}^{+0.03}$  &  $0.80_{-0.01}^{+0.02}$ && $24\pm2$& $4160.00/3742$    \\ \hline \hline
       & $kT$  & $kT_{\textup{mid}}$  & $kT_{\textup{low}}$ & $\textup{Norm}/\textup{Norm}_{\textup{mid}}$ & $\textup{Norm}_{\textup{mid}}/\textup{Norm}_{\textup{low}}$ &  C-stat/d.o.f. & AIC\\
   & (keV) &  (keV) & (keV) &  &\\    \hline
    Resolve                &             &           &             &            \\
    1T                &  $2.37_{-0.02}^{+0.02}$   & -- & -- &--&--& $11997.01/12878$ & 12023.01  \\
    1T w/ zgauss     &    $2.39_{-0.02}^{+0.02}$  & -- & -- &--&--& $11973.12/12876$  & 12003.12 \\
    2T                &    $3.02_{-0.11}^{+0.12}$ & $1.63_{-0.09}^{+0.08}$  &  --  &$1.0\pm0.2$&--& $11839.35/12876$  &11869.35\\
    2T w/ zgauss     & $3.02_{-0.11}^{+0.09}$ & $1.54_{-0.11}^{+0.07}$ &  -- &$1.2\pm0.3$&--& $11778.75/12874$  & 11812.75\\ 
   3T w/ zgauss  &  $3.05_{-0.12}^{+0.08}$ & $1.59_{-0.10}^{+0.07}$ & $0.80$ (fix)  & $1.1\pm0.3$ &24 (fix)& $11776.64/12874$ & 11810.64\\
    \hline \hline
   & $kT_\textup{mean}$  & $\sigma_{kT}$ & $kT_{\textup{low}}$ & && C-stat/d.o.f. &AIC \\
   & (keV) & (keV) & (keV)  & &\\    \hline
    Resolve                &             &           &             &            \\
    gadem w/ zgauss  & $2.25_{-0.06}^{+0.04}$ & $1.22_{-0.06}^{+0.10}$ & -- &&& $11777.26/12875$ & 11809.26        \\
    gadem+1T w/ zgauss  & $2.29_{-0.10}^{+0.09}$ & $1.17_{-0.11}^{+0.15}$  & 0.80 (fix) &  && $11776.65/12874$    &  11810.65     \\
    \hline
  \end{tabular} } \label{tab:fit_params}
\begin{tabnote} 
\end{tabnote}
\end{table*}

\begin{table}
  \tbl{Comparison of the resultant Fe, bulk velocity, and velocity dispersion with each model.}{%
  \begin{tabular}{lccc}
    \hline
 Model  & Fe & & Velocity Dispersion   \\
    & (Solar) & & (km\,s$^{-1}$) \\    \hline 
        RGS            &             &     \\
   2T       & $1.76_{-0.05}^{+0.09}$ &  & 120 (fix)   \\ \hline \hline
   & Fe & Bulk Velocity$^\ast$ & Velocity Dispersion \\
   &  (Solar) & (km\,s$^{-1}$) & (km\,s$^{-1}$)  \\    \hline
    Resolve                &             & &                      \\
    1T                   &  $1.62_{-0.04}^{+0.04}$ & $-163_{-4}^{+4}$ & $124_{-5}^{+5}$      \\
    1T w/ zgauss         & $1.70_{-0.04}^{+0.04}$ & $-165_{-4}^{+4}$ & $120_{-7}^{+7}$    \\
    2T                     & $1.60_{-0.04}^{+0.04}$ & $-161_{-4}^{+4}$& $126_{-5}^{+5}$  \\
    2T w/ zgauss        & $1.73_{-0.04}^{+0.04}$ & $-165_{-4}^{+4}$ & $122_{-7}^{+7}$  \\ 
   3T w/ zgauss   & $1.73_{-0.04}^{+0.04}$ & $-165_{-4}^{+4}$ & $122_{-7}^{+7}$\\
    \hline \hline
   &  Fe & Bulk Velocity & Velocity Dispersion  \\
   & (Solar) & (km\,s$^{-1}$) & (km\,s$^{-1}$)  \\    \hline
    Resolve                &             &           &                        \\
    gadem w/ zgauss   & $1.74_{-0.04}^{+0.04}$ & $-165_{-4}^{+4}$& $122_{-7}^{+7}$          \\
    gadem+1T w/ zgauss &  $1.74_{-0.05}^{+0.04}$ & $-165_{-4}^{+4}$ & $122_{-7}^{+7}$    \\
    \hline
  \end{tabular}} \label{tab:fit_params2}
\begin{tabnote} 
\footnotemark[$\ast$] Line-of-sight bulk velocity relative to NGC~4696, which has $3008 \pm 7$\,km\,s$^{-1}$ of the heliocentric velocity.\\
\end{tabnote}
\end{table}

\section{Observation and data reduction}

\subsection{XRISM/Resolve}
XRISM carried out an observation of the core region of the Centaurus cluster from December 28th, 2023, to January 3rd, 2024 (OBSID=000138000). The aim point was located $1'$ northwest of NGC~4696. We used only the Resolve instrument \citep{Kelley et al. JATIS 2025} in this paper. The observed region with XRISM/Resolve is shown in figure \ref{fig:img_RGS}. During this observation, the gate valve was closed, and data in the $E \gtrsim$2 keV band were available. Data reduction was basically performed on the entire field of view (FoV) of Resolve according to the XRISM Data Reduction Guide\footnote{https://heasarc.gsfc.nasa.gov/docs/xrism/analysis/abc\_guide/xrism\_abc.html}, except that pixel 27 was included in the spectral analysis as described in \citet{2025Natur.638..365X}. 
The non-X-ray background (NXB) model\footnote{https://heasarc.gsfc.nasa.gov/docs/xrism/analysis/nxb/nxb\_spectral\_models.html} of the Resolve  distributed by the XRISM team was included within the spectral fit model.
The cleaned exposure time after screening was 287.4 ksec. We used only high-resolution primary events. In this study, we only performed the spectral fitting from the entire region of the Resolve, without dividing the region into subregions.

\subsection{XMM-Newton/RGS} \label{sect:XMM-Newton/RGS}

Two XMM-Newton observations capture the center of NGC 4696 at the RGS pointing position \citep{2001A&A...365L...7D}. In this paper, we used only the longer exposure data (obsID$=$0406200101) to avoid the impact of differences in roll angles and observation coordinates on the fitting results, as shown in \citet{Mernier2025submitted}. The data reduction procedure of the XMM-Newton RGS data was performed with XMM-Newton Science Analysis System (SAS) version 21.0. We processed the RGS1 and RGS2 data with the {\tt rgsproc} task in the SAS package. 
The exposure times after filtering for RGS1 and RGS2 were 119.2 ksec and 113.8 ksec, respectively. Following \citet{2022MNRAS.514.4222F}, the first- and second-order RGS spectra were extracted by specifying the rectangular cross-dispersion limits of the source regions, rather than using the {\tt xpsfincl} task in the SAS package. We extracted the spectra with a 2$'$ width region along the cross-dispersion direction within the central chip of the MOS detectors, centered on (R.A., Dec.)$_\textup{J2000.0}=$(\timeform{12h48m49.05s}, \timeform{-41D18'46.1''}). This spectral extraction region is shown in figure \ref{fig:img_RGS}.
As for the dispersion direction, it is difficult to derive the spatial information from the RGS spectra of extended sources, such as the ICM.

\section{Spectral analysis and results}

\subsection{Spectral analysis with XRISM/Resolve} \label{subsection:Thermal models}
\subsubsection{Single-temperature approach}

We first fitted the Resolve spectrum for the entire region in the 1.8--12 keV energy band, with a thermal plasma model (bvvapec model, \cite{2001ApJ...556L..91S}), multiplied by an absorption model (tbabs model, \cite{2000ApJ...542..914W}) to account for the Galactic absorption. The spectrum is binned so as to have a minimum of 1 count per energy bin. We fit the spectrum using C-statistics \citep{1979ApJ...228..939C}.
The model formula can be expressed as follows: $tbabs\times bvvapec$ (hereafter, we refer to this model as the ``1T'' model). We fixed the abundances of elements from C to Al at 1.5 Solar. The abundances of Si, S, Ar, Ca, Cr, Mn, Fe, and Ni were left free, while those of the remaining elements between Si and Zn were linked to the Fe abundance.

The spectrum is shown in figure \ref{fig:spec_Resolve1T2T} and the best-fit parameters are summarized in tables \ref{tab:fit_params} and \ref{tab:fit_params2}. 
Although the overall continuum is reasonably reproduced with the ``1T'' model, positive residuals are seen at the S He$\alpha$ lines and at the Fe Ly$\alpha$ lines, while negative residuals appear at the Fe \emissiontype{XXV} He$\alpha$ resonance ($w$) line (see figure \ref{fig:spec_Resolve1T2T}). 
If this resonance line is suppressed by resonant scattering, the velocity dispersion would be overestimated, whereas the Fe abundance would be underestimated, because both quantities
are predominantly constrained by the line profile of the Fe He$\alpha$ complex \citep{2018PASJ...70...10H}.
 To mitigate this issue, we excluded the Fe \emissiontype{XXV} He$\alpha$ $w$ line from \atomdb{} and added a $zgauss$ component into the model formula to the line (``Resolve 1T w/ zgauss'' model). 
The line centroid energy for the $zgauss$ component was fixed at $6.7004$ keV\@, and its redshift was tied to that of the ICM ($bvvapec$). Tables \ref{tab:fit_params} and \ref{tab:fit_params2} summarizes the resultant fit parameters. As shown in figure \ref{fig:spec_Resolve1T2T}, the fit around Fe \emissiontype{XXV} He$\alpha$, particularly at the $w$ line, improved significantly. The velocity dispersion becomes slightly smaller, and the Fe abundance becomes slightly larger than those with the original ``1T'' model, although the change in velocity dispersion remains within the $1\sigma$ uncertainty. This behavior is consistent with the possibility that resonant scattering is suppressing the $w$ line in the Centaurus cluster core.

\subsubsection{Two-temperature approach}

\citet{2025Natur.638..365X} reported that the Resolve spectrum of the central $1.5'\times 2.0'$ region in the sub-array analysis required two temperature components. We therefore added an additional thermal component, expressed as follows: $tbabs\times (bvvapec+bvvapec)$ (hereafter the Resolve ``2T'' model), and fitted the Resolve full-array spectrum. The redshift, velocity dispersion, and abundance of each component were tied and treated as free parameters. 
To model the Fe \emissiontype{XXV} He$\alpha$ $w$ line in the same manner as in the Resolve ``1T w/ zgauss'' fit, 
we also defined a model of the form $tbabs\times (bvvapec+bvvapec+zgauss)$, referred to as the ``2T w/ zgauss'' model. Note that this model includes only one $zgauss$ component to represent the Fe \emissiontype{XXV} He$\alpha$ resonance line. Hereafter, we denote the two temperatures derived from both the ``2T'' and ``2T w/ zgauss'' models as $kT$ and $kT_\textup{mid}$. 
The resultant parameters of these two fits are listed in tables \ref{tab:fit_params} and \ref{tab:fit_params2}.
As shown in lower panels of figure \ref{fig:spec_Resolve1T2T}, the ``2T'' model provides a better fit to the S He$\alpha$ lines and to the Fe Ly$\alpha$ lines, which show positive residuals in the 1T fit. This improvement leads to the lower C-statistic value obtained with the ``2T'' model.
The ``2T w/ zgauss'' model further improves the fit around the Fe \emissiontype{XXV} He$\alpha$ complex, particularly at the resonance line, and yields a lower C-statistic value.
The Fe abundance, bulk velocity, and velocity dispersion obtained with the ``2T w/ zgauss'' model are close to those derived from the ``1T w/ zgauss'' model. 
We further evaluated the models with the Akaike Information Criterion (AIC; \cite{1974ITAC...19..716A}) to compare their goodness of fits. The AIC is defined as $\mathrm{AIC}_i =  - 2 \ln(L_i) + 2k_i$, where $L_i$ is the maximized likelihood function, $k_i$ is the number of free parameters of model $i$. The smaller the AIC value, the better the model fits. By comparing the AIC values of ``1T w/ zgauss'' model and ``2T w/ zgauss'' model, $\Delta \mathrm{AIC} = 190.37$ was determined. Similarly, by comparing ``2T'' model and ``2T w/ zgauss'' model, $\Delta \mathrm{AIC} = 56.60$ was also calculated. Thus, ``2T w/ zgauss'' model provides a superior description of the spectrum compared with ``1T w/ zgauss'' model, and model ``2T w/ zgauss'' provides a superior description compared with model ``2T''. 
We also tested models with an additional temperature component, but the fits did not improve, and the temperature of the new component could not be constrained.

\subsubsection{DEM approach}

We also fitted a multi-temperature model with a differential emission measure (DEM) to the Resolve spectrum
expressed as $tbabs\times(bvvgadem+zgauss)$, hereafter ``gadem w/ zgauss''.
In the $bvvgadem$ model, the emission measure follows a Gaussian distribution characterized by a mean temperature,  $kT_\textup{mean}$,  and a standard deviation, $\sigma_{kT}$.
As in the ``1T w/ zgauss'' and ``2T w/ zgauss'' models, we excluded the Fe \emissiontype{XXV} He$\alpha$ $w$ line from \atomdb{} and represented it using a $zgauss$ model (hereafter ``gadem w/ zgauss'').  The resultant fit parameters are listed in tables \ref{tab:fit_params} and \ref{tab:fit_params2}. The derived mean temperature is close to that from the ``1T'' model, while $\sigma_{kT}$
is comparable to the separation of the two temperatures derived from the ``2T w/ zgauss'' model. 
The C-statistic value, Fe abundance, bulk velocity, and velocity dispersion are nearly identical to those of the ``2T w/ zgauss'' model. As shown in lower panels of figure \ref{fig:spec_Resolve1T2T}, the 2T and $gadem$ models had no significant deviation in the residual distribution.

\begin{figure}[t]
\begin{center}
\includegraphics[width=80mm]{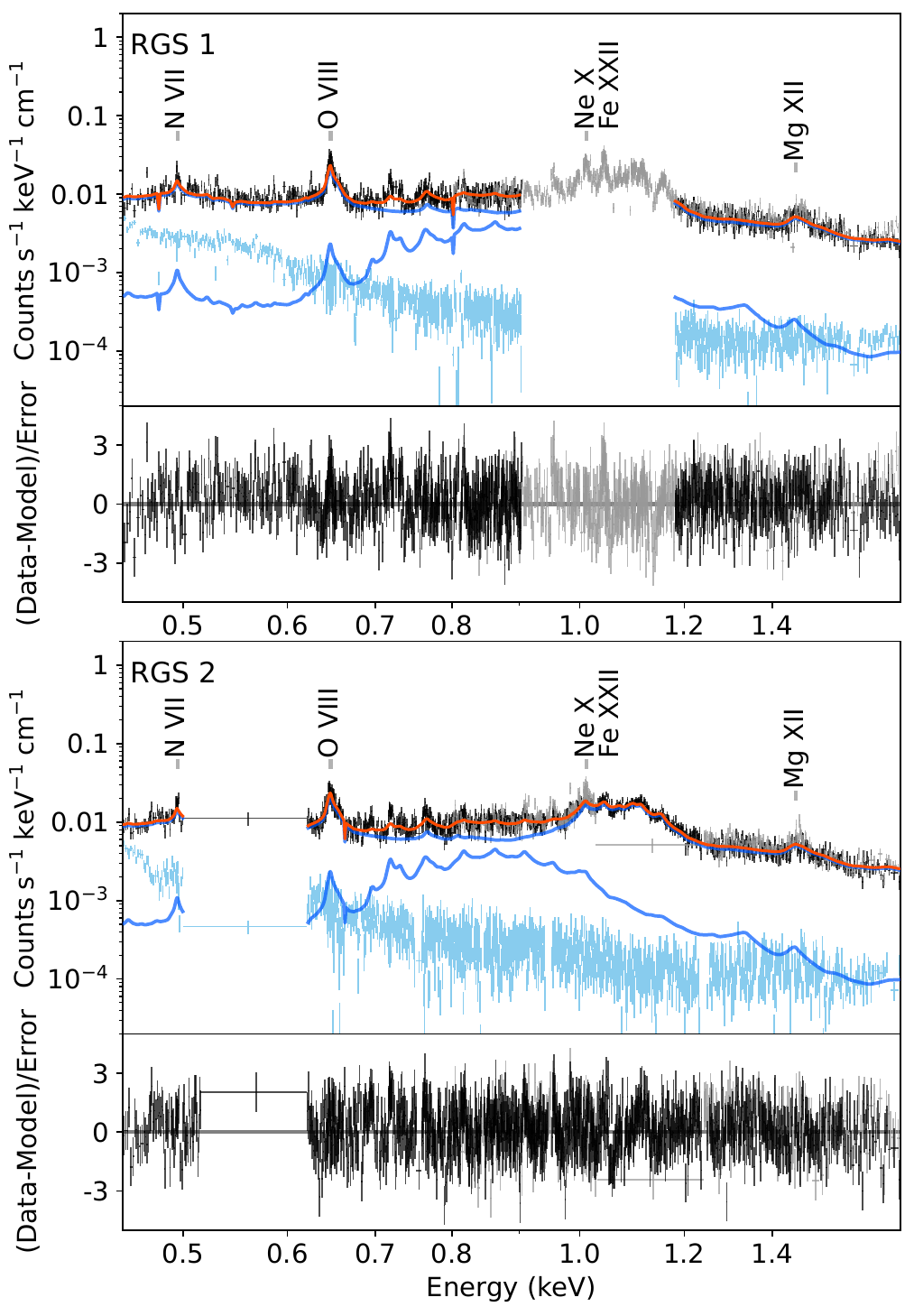}
\end{center}
\caption{The RGS spectra of the Centaurus cluster with a 2 arcmin extraction width. The black data points represent the first-order spectra, and the light gray data points represent the second-order spectra. The 0.90--1.18 keV and 0.52--0.62 keV bands of RGS1 and RGS2 are unavailable because CCD7 of RGS1 and CCD4 of RGS2 are unavailable. The red solid line represents the best-fit model. The blue lines indicate the two-temperature components. The models are plotted only where CCDs are working. The light blue data points represent the background spectra. The lower panels display the residuals between the data and the model. {Alt text: Two line graphs. In the upper panel of each graph, the y-axis shows the count rate per second per kilo–electron volt. In the lower panel of each graph, the y-axis shows the residuals between the data and the model. The x-axis in all panels shows the energy.} } 
\label{fig:spec_RGS2T}
\end{figure}

\begin{figure*}[t]
\begin{minipage}{1.0\hsize}
\begin{center}
\includegraphics[width=160mm]{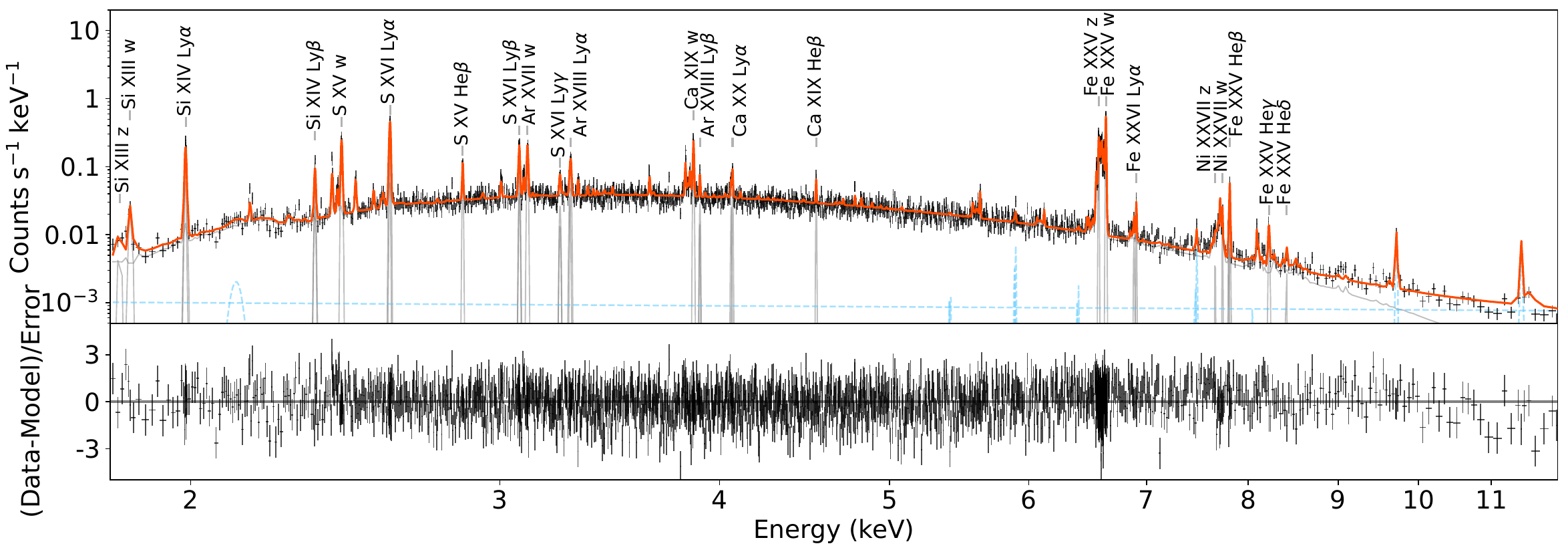}
\end{center}
\end{minipage}
\caption{Resolve spectrum of the Centaurus cluster. The red solid line represents the best-fit model described in section \ref{sec:Line ratio diagnostics}, in which the lines listed in table \ref{tab:line_list} are excluded from the $bvvapec$ model and $zgauss$ components are added to the Resolve ``1T w/ zgauss'' model. The light gray lines show the model of $zgauss$ components. The light blue dashed line represents the NXB model. The lower panel displays the residuals between the data and the model. {Alt text: One line graph. In the upper panel, the y-axis shows the count rate per second per kilo–electron volt. In the lower panel, the y-axis shows the residuals between the data and the model. The x-axis shows the energy.}} 
\label{fig:spec_Resolve1T_gauss}
\end{figure*}

\begin{figure*}[!th]
\begin{minipage}{1.0\hsize}
\begin{center}
\includegraphics[width=145mm]{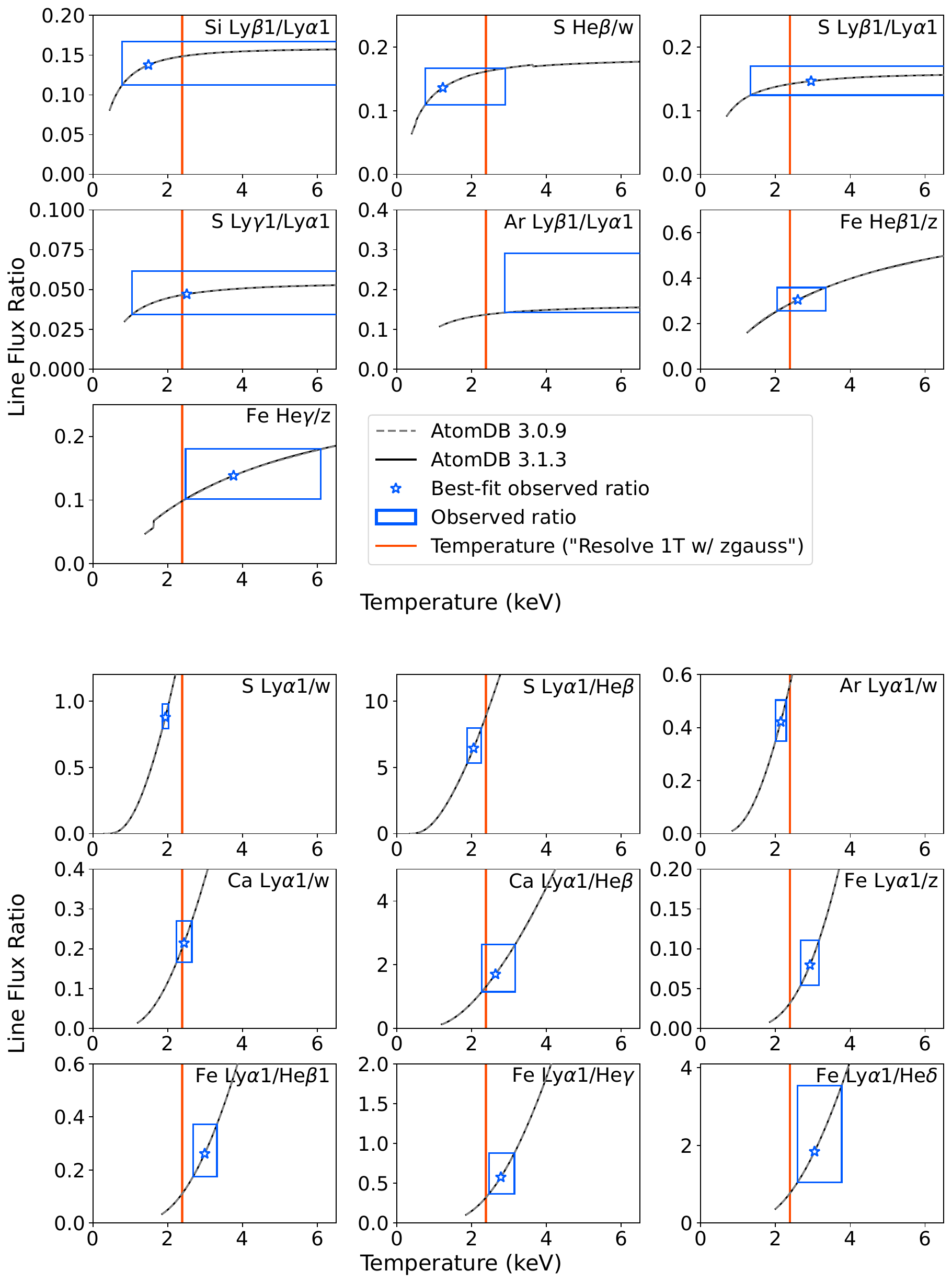}
\end{center}
\end{minipage}
\caption{Upper seven panels show the line flux ratio against the excitation temperatures. Gray dashed lines and black solid lines represent the temperature dependence of the line flux ratios based on \atomdb{} 3.0.9 and 3.1.3, respectively. The star-shaped markers show the best-fit observed ratios. The blue boxes indicate the range of observed ratios and the corresponding excitation temperatures derived from \atomdb{} 3.1.3. The red bands represent the temperature derived from the Resolve ``1T w/ zgauss'' model. The lower nine panels follow the same format, but the x-axis corresponds to the ionization temperatures. {Alt text: Sixteen line-ratio plots comparing temperatures derived from the observed data with those predicted by AtomDB 3.0.9 and 3.1.3. Blue squares indicate the best-fit results.} } 
\label{fig:line_ratio}
\end{figure*}

\subsection{Joint analysis with XMM-Newton/RGS}

\subsubsection{Constraint of the cool component with the RGS spectra}
\label{subsec:rgs}

Previous studies have reported the presence of an X-ray-emitting component cooler than 1 keV in the Centaurus cluster core (e.g., \cite{2008MNRAS.385.1186S, 2009ApJ...701..377T, 2022MNRAS.514.4222F}). The Resolve instrument has no sensitivity below the 2 keV bandpass because the gate valve is closed and therefore cannot constrain the emission of such a cool gas component. 
We analyzed the RGS data to assess this cool-phase gas and then incorporated it into the Resolve spectral analysis.
In this work, we adopted a $2'$-width extraction region for RGS, 
which is smaller than the Resolve FoV as shown in figure \ref{fig:img_RGS}.
Since the cooler component below 1 keV 
would be more centrally concentrated than the hot-phase ICM \citep{2015A&A...575A..38P}, the $2'$ RGS extraction region should still contain most of this compact cool component. 
In addition, widening the RGS extraction region would degrade the effective spectral resolution and increase the contribution of the hotter gas, making it less suitable for detecting the cool component.
 For these reasons, we used the 2$'$-wide RGS data to constrain the low-temperature component in the Resolve spectral modeling.

The RGS spectra were grouped so that each bin had at least 20 counts, and fitted by minimizing the $\chi^{2}$ value. The first- and second-order spectra of RGS were fitted in the 0.45--1.75 and 0.80--1.75 keV energy bands, respectively, as shown in figure \ref{fig:spec_RGS2T}. The RGS1 and RGS2 spectra, including both the first- and second-order spectra, were fitted simultaneously. 
To account for the spectral broadening caused by the spatial extent of the source, we used the $rgsxsrc$ model in XSPEC with the MOS image in the 0.45--2.0 keV band. 
We generated the background spectra using the SAS tool {\tt rgsbkgmodel} and subtracted them from the source spectra. Similar to the spectral fits of the Resolve, the RGS spectra were fitted with the two-temperature model, multiplied by an absorption model. The model formula is expressed as follows: $rgsxsrc \times tbabs\times (bvvapec+bvvapec)$ (``RGS 2T'' model). The velocity dispersion was fixed at $120$\,km\,s$^{-1}$ in the ``RGS 2T'' model. We treated the Galactic hydrogen column density $N_{\textup{H}}$ as a free parameter. The resultant parameters are shown in tables \ref{tab:fit_params} and \ref{tab:fit_params2}. 
The RGS analysis provides two temperature components, 1.9 and 0.8 keV. The higher-temperature component is consistent with $kT_\textup{mid}$ ($\sim 1.6$\,keV) found with the Resolve ``2T w/ zgauss'' model, implying that they represent the same phase gas. The lower-temperature component ($kT_\textup{low}$) matched the interstellar medium associated with NGC~4696 \citep{2009ApJ...701..377T} 
and is also consistent with the central temperature measured with XMM-Newton \citep{2016MNRAS.457...82S}, 
supporting the assumption of the localized cooler component.

We investigated the dependence of the derived temperatures on the assumed velocity dispersion by fixing it at $1000$\,km\,s$^{-1}$ or allowing it to vary freely.
Both cases yielded consistent temperatures with the value in table\,\ref{tab:fit_params} within the statistical uncertainties.
To examine the impact of the RGS background subtraction method, we performed, in addition to the subtraction of the RGS background spectrum, two alternative tests: (i) representing the RGS background spectrum by a simple power-law component and (ii) modelling it with a more complex formula ($constant\times ( bknpow+ gauss+gauss) $). In all cases, the temperatures derived from the ``RGS 2T'' model fits are consistent. 
We tested three configurations of $N_{\textup{H}}$ in which it was fixed at $7.77\times 10^{20}$\,cm$^{-2}$ as adopted in the Resolve spectral fits, treated as a free parameter, and fixed at $1.22\times 10^{21}$\,cm$^{-2}$ \citep{2013MNRAS.431..394W}.
In all cases, the derived temperatures are consistent with each other. The two-temperature model applied by \citet{2008MNRAS.385.1186S} yielded temperatures of $1.85\pm0.02$ keV and $0.766\pm0.008$ keV which are in good agreement with our results.
Like the temperatures, the RGS and XRISM Fe abundances reported here are also fairly consistent with each other, although we note more generally that the Fe abundances from the RGS analysis are sensitive to many modelling assumptions \citep{2017A&A...607A..98D,2022MNRAS.514.4222F}, and a similar analysis of the same RGS data in \citet{Mernier2025submitted} yields considerably different Fe abundances.

\begin{table*}
  \tbl{List of the emission lines removed in the $bvapec$ model and the resultant fit parameters of the ``zgauss''  model for line diagnostics. }{%
  \begin{tabular}{ccccccc}
      \hline
      Line name & E & \multicolumn{3}{c}{Constraints} & Flux & Width \\ \cline{3-5}
       & (keV) & Tied to & Width & Flux & (photons\,cm$^{-2}$\,s$^{-1}$) & (eV)\\       \hline 
Si \emissiontype{XIII} $z$ & 1.8394 & Si \emissiontype{XIII} w & $\times 1$ & -- & $1.4_{-0.5}^{+0.5}\times10^{-4}$ & -- \\   
Si \emissiontype{XIII} $w$ & 1.8650 & -- & -- & -- & $2.8_{-0.5}^{+0.6}\times10^{-4}$&$2.4_{-0.8}^{+1.1}$\\   
Si \emissiontype{XIV} Ly$\alpha_1$ & 2.0061 & -- & -- & -- & $4.0_{-0.3}^{+0.3} \times10^{-4}$ &$1.2_{-0.3}^{+0.3}$\\
Si \emissiontype{XIV} Ly$\alpha_2$ & 2.0043 & Si \emissiontype{XIV} Ly$\alpha_1$ & $\times 1$ & $\times 0.5$ & -- & -- \\
Si \emissiontype{XIV} Ly$\beta_1$ & 2.3766 & -- & -- & --& $5.5_{-0.7}^{+0.7} \times10^{-5}$ 
 & $1.2_{-0.6}^{+0.5}$\\
Si \emissiontype{XIV} Ly$\beta_2$ & 2.3761 & Si \emissiontype{XIV} Ly$\alpha_2$ & $\times 1$ & $\times 0.5$ & -- & -- \\
S \emissiontype{XV} $w$ & 2.4606 & -- & -- & -- & $1.9_{-0.2}^{+0.2} \times10^{-4}$ & $1.5_{-0.3}^{+0.3}$ \\
S \emissiontype{XV} He$\beta_1$ \footnotemark[*]& 2.8840 & -- & -- & -- & $2.6_{-0.4}^{+0.4} \times10^{-5}$& $<1.3$\\
S \emissiontype{XVI} Ly$\alpha_1$ & 2.6227 & -- & -- & -- & $1.7_{-0.1}^{+0.1} \times10^{-4}$ & $1.0_{-0.3}^{+0.3}$\\
S \emissiontype{XVI} Ly$\alpha_2$ & 2.6197 & S \emissiontype{XVI} Ly$\alpha_1$ & $\times 1$ & $\times 0.5$ & -- & -- \\
S \emissiontype{XVI} Ly$\beta_1$ & 3.1067 & -- & -- & -- & $2.5_{-0.3}^{+0.3} \times10^{-5}$& $1.8_{-0.4}^{+0.4}$\\
S \emissiontype{XVI} Ly$\beta_2$ & 3.1059 & S \emissiontype{XVI} Ly$\beta_1$ & $\times 1$ & $\times 0.5$ & -- & -- \\
S \emissiontype{XVI} Ly$\gamma_1$ & 3.2763 & -- & -- & -- & $7.9_{-1.5}^{+1.6} \times10^{-6}$& $2.3_{-0.8}^{+0.8}$\\
S \emissiontype{XVI} Ly$\gamma_2$ & 3.2759 & S \emissiontype{XVI} Ly$\gamma_1$ & $\times 1$ & $\times 0.5$ & -- & -- \\
Ar \emissiontype{XVII} $w$ & 3.1396 & -- & -- & -- &$4.7_{-0.4}^{+0.4} \times10^{-5}$ & $2.0_{-0.3}^{+0.3}$\\
Ar \emissiontype{XVIII} Ly$\alpha_1$ & 3.3230 & -- & -- & -- & $2.0_{-0.2}^{+0.3} \times10^{-5}$& $2.2_{-0.6}^{+0.7}$\\
Ar \emissiontype{XVIII} Ly$\alpha_2$ & 3.3182 & Ar \emissiontype{XVIII} Ly$\alpha_1$ & $\times 1$ & $\times 0.5$ & -- & -- \\
Ar \emissiontype{XVIII} Ly$\beta_1$ & 3.9357 & -- & -- & -- & $4.2_{-0.9}^{+1.0} \times10^{-6}$& $<2.3$\\
Ar \emissiontype{XVIII} Ly$\beta_2$ & 3.9343 & Ar \emissiontype{XVIII} Ly$\beta_1$ & $\times 1$ & $\times 0.5$ & -- & -- \\
Ca \emissiontype{XIX} $w$ & 3.9024 & -- & -- & -- & $3.3_{-0.3}^{+0.3} \times10^{-5}$ & $2.0_{-0.3}^{+0.3}$\\
Ca \emissiontype{XIX} He$\beta_1$ \footnotemark[*] & 4.5835 & -- & -- & -- & $4.2_{-1.0}^{+1.0} \times10^{-6}$& $<2.2$\\
Ca \emissiontype{XX} Ly$\alpha_1$ & 4.1075 & -- & -- & -- &$7.1_{-1.1}^{+1.1} \times10^{-6}$& $1.4_{-0.9}^{+0.8}$\\
Ca \emissiontype{XX} Ly$\alpha_2$ & 4.1001 & Ca \emissiontype{XX} Ly$\alpha_1$ & $\times 1$ & $\times 0.5$ & -- & -- \\
Fe \emissiontype{XXV} $z$ & 6.6366 & -- & -- & -- & $3.0_{-0.2}^{+0.2} \times10^{-5}$ & $2.4_{-0.3}^{+0.3}$\\
Fe \emissiontype{XXV} $w$ & 6.7004 & -- & -- & -- & $6.5_{-0.3}^{+0.3} \times10^{-5}$& $3.1_{-0.2}^{+0.2}$\\
Fe \emissiontype{XXV} He$\beta_1$ & 7.8815 & -- & -- & -- & $9.2_{-0.9}^{+1.0} \times10^{-6}$ & $3.0_{-0.4}^{+0.5}$\\
Fe \emissiontype{XXV} He$\beta_2$ & 7.8720 & Fe \emissiontype{XXV} He$\beta_1$ & $\times 1$ & -- & $2.3_{-0.6}^{+0.7} \times10^{-6}$ & --\\
Fe \emissiontype{XXV} He$\gamma_1$ \footnotemark[*] & 8.2955 & -- & -- & -- & $4.2_{-0.8}^{+0.8} \times10^{-6}$ &$6.1_{-1.4}^{+1.6}$\\
Fe \emissiontype{XXV} He$\delta_1$ \footnotemark[*]& 8.4874 & -- & -- & -- & $1.3_{-0.5}^{+0.5} \times10^{-6}$ &$1.9_{-1.2}^{+1.2}$\\
Fe \emissiontype{XXVI} Ly$\alpha_1$ & 6.9731 & -- & -- & -- & $2.4_{-0.6}^{+0.6} \times10^{-6}$ &$1.6_{-1.5}^{+1.2}$\\
Fe \emissiontype{XXVI} Ly$\alpha_2$ & 6.9519 & -- & -- & -- & $1.6_{-0.5}^{+0.6} \times10^{-6}$ &$3.1_{-1.4}^{+1.7}$\\
Ni \emissiontype{XXVII} $z$ & 7.7316 & Ni \emissiontype{XXVII} w & $\times 1$ & -- & $9.2_{-3.9}^{+4.6} \times10^{-7}$ &--\\
Ni \emissiontype{XXVII} $w$ & 7.8056 & -- & -- & -- & $2.2_{-0.6}^{+0.6} \times10^{-6}$ & $<2.4$\\
      \hline
    \end{tabular}}\label{tab:line_list}
\begin{tabnote}
\footnotemark[$\ast$] Each complex (S \emissiontype{XV} He$\beta$, Ca \emissiontype{XIX} He$\beta$, Fe \emissiontype{XXV} He$\gamma$, and Fe \emissiontype{XXV} He$\delta$) was represented by a single $zgauss$ component.
\end{tabnote}
\end{table*}

\begin{figure}[t]
\begin{center}
\includegraphics[width=80mm]{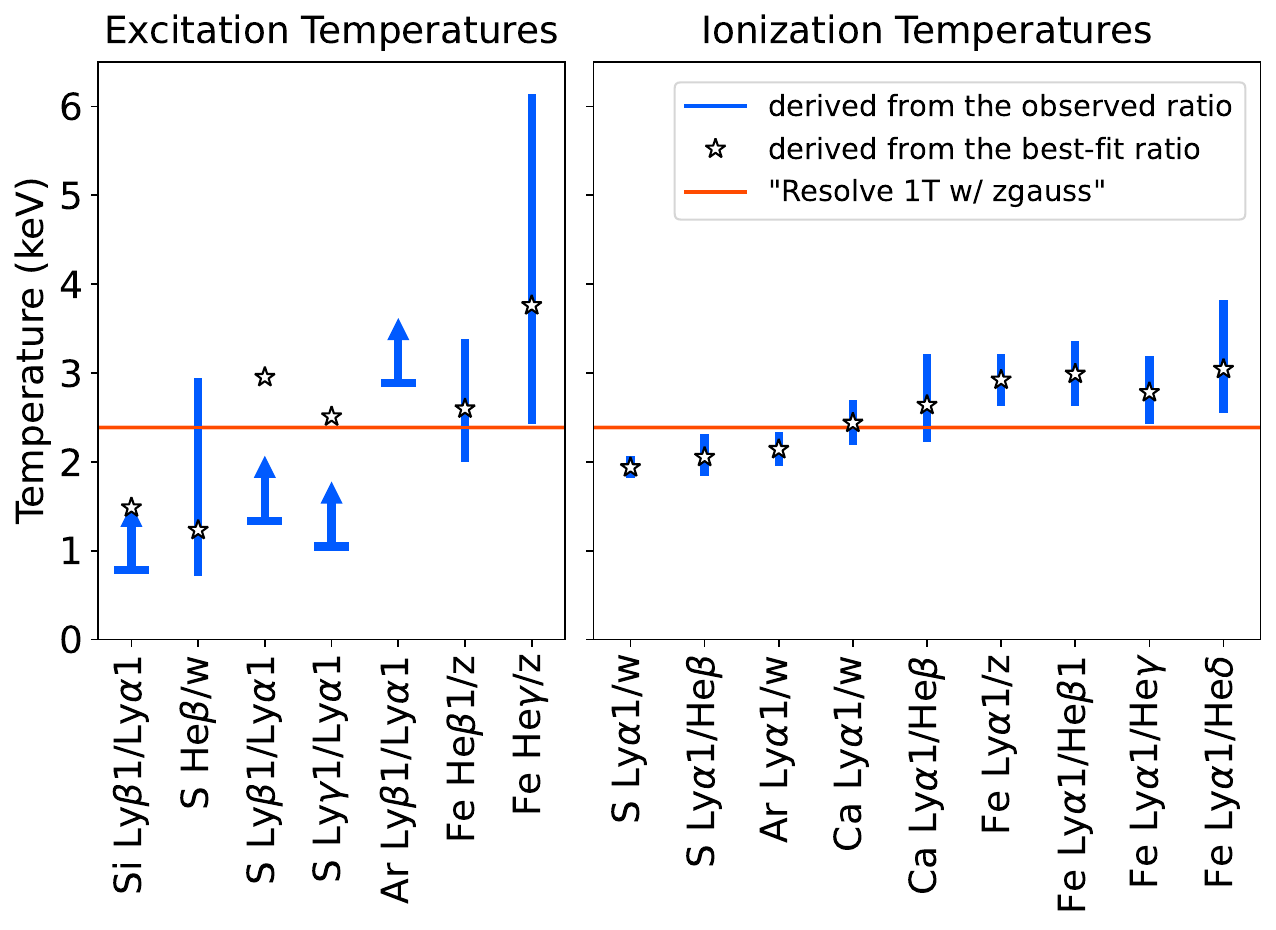}
\end{center}
\caption{Summary of the derived excitation and ionization temperatures from line flux ratios. The star-shaped markers show the excitation and ionization temperatures from the resultant best-fit line flux ratios. The horizontal red line shows the temperature from the Resolve ``1T w/ zgauss'' model. {Alt text: Two line graphs. The y-axis of each panel shows the temperature from 0 keV to 6.5 keV. Each data point represents the temperature derived from an individual line flux ratio.}}
\label{fig:kT_line_ratio}
\end{figure}

\begin{figure}[t]
\begin{center}
\includegraphics[width=80mm]{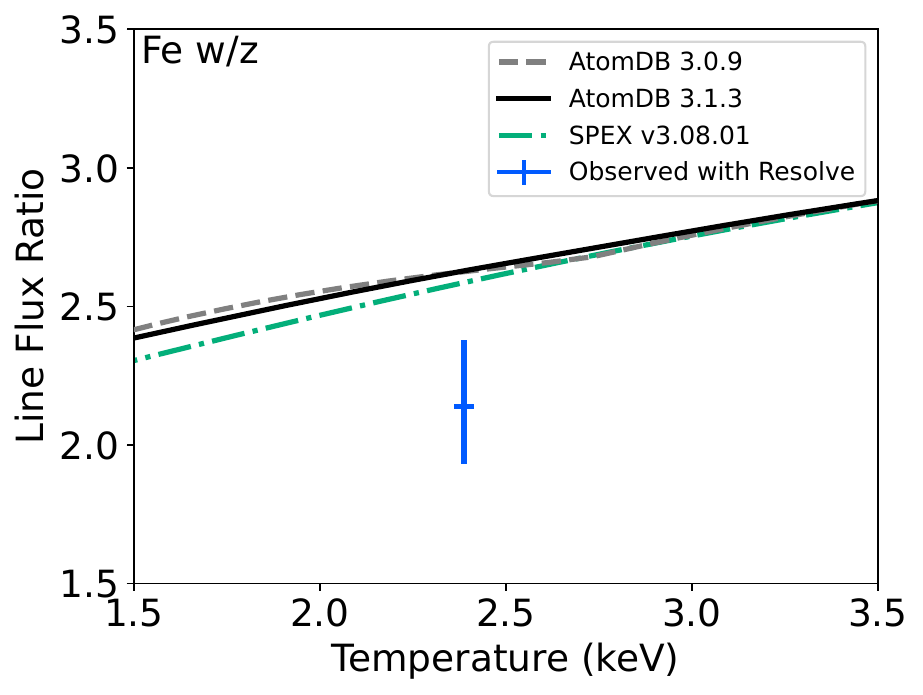}
\end{center}
\caption{Comparisons of the observed Fe \emissiontype{XXV} He$\alpha$ $w/z$ line ratio with model predictions with \atomdb{} 3.0.9, 3.1.3, and SPEX v3.08.01. The vertical error bar of the blue cross represents the observed $w/z$ ratio, with its horizontal error corresponding to the temperature uncertainty derived from the Resolve ``1T w/ zgauss'' model. The gray dashed line, black solid line, and green dash-dotted line indicate the temperature dependences of the ratio for \atomdb{} 3.0.9, 3.1.3, and SPEX v3.08.01, respectively. {Alt text: One line graphs. The y-axis shows the line flux ratio. The x-axis show the temperature from 1.5 keV to 3.5 keV.}} 
\label{fig:kT_Fe_wtoz}
\end{figure}

\subsubsection{Impact on the spectral analysis for Resolve}

Since the $kT_{\textup{low}}$ component could not be constrained from the Resolve spectral fits alone, we included the additional $kT_{\textup{low}}$ component from the RGS analysis into the Resolve analysis. We fixed the temperature of the $kT_{\textup{low}}$ component and also fixed the normalization ratio of the $kT_\textup{mid}$ to $kT_\textup{low}$ components derived from the RGS spectral fit. The model formula is represented as $tbabs\times (bvvapec+bvvapec+bvvapec+zgauss)$ (Resolve ``3T w/ zgauss'' model). 
Although the spatial extraction regions of RGS and Resolve are not identical, we tested the effect of including the RGS-derived low-temperature component using the RGS normalization ratio as a trial assumption.
The resultant parameters are shown in tables \ref{tab:fit_params} and \ref{tab:fit_params2}. The parameters derived from the Resolve ``2T w/ zgauss'' model are consistent with those from the ``3T w/ zgauss'' model.

We also applied the same approach as in the Resolve ``3T w/ zgauss'' model to the ``gadem w/ zgauss'' model. We added the $kT_\textup{low}$ components derived from the RGS analysis into the DEM model. The model formula is $tbabs\times (bvvgadem+bvvapec+zgauss)$ (``gadem+1T w/ zgauss'' model). The temperature of the $bvvapec$ model was fixed to the $kT_\textup{low}$ value from the RGS analysis. The resultant best-fit parameters are listed in tables \ref{tab:fit_params} and \ref{tab:fit_params2}. The temperatures from the ``gadem+1T w/ zgauss'' model (with $kT_\textup{low}$ component) and the ``gadem w/ zgauss'' model (without $kT_\textup{low}$ component) are consistent with each other. Thus, the $kT_\textup{low}$ component would have no impact on the results of the Resolve spectral fits above 2 keV, regardless of whether the main band is reproduced by the two-temperature or DEM approach.

\subsection{Line ratio diagnostics} \label{sec:Line ratio diagnostics}

As shown in figure \ref{fig:spec_Resolve1T_gauss}, emission lines from Si to Ni are clearly detected.
While the continuum of the Resolve spectrum, whose shape roughly represents the electron temperature, is well represented by both the ``1T'' and ``2T'' approaches,
the ``1T'' model fails to explain the intensities for several lines (see figure \ref{fig:spec_Resolve1T2T}).
Due to the strong dependence of line emissivities on plasma temperature, detailed diagnostics of observed line fluxes serve as a sensitive probe of the temperature structure.

To measure the fluxes of the individual lines, we excluded those listed in table \ref{tab:line_list} from the $bvvapec$ model, and added $zgauss$ components to the Resolve ``1T w/ zgauss model''. 
The rest-frame centroid energy of each line in table \ref{tab:line_list} was taken from \atomdb{} 3.1.3. Its redshift was tied to that of the $bvvapec$ model and left free to vary.  
We adopted the 1T model for this analysis because it already reproduces the observed continuum well, and as shown in figure \ref{fig:spec_Resolve1T_gauss}, the continuum is accurately modeled after adding the Gaussian components. Therefore, an additional temperature component is unnecessary for the purpose of measuring individual line fluxes.
Emission lines not listed in table \ref{tab:line_list} remained included in the $bvvapec$ model.  The abundance and velocity dispersion of the $bvvapec$ model were fixed to the numbers derived from the Resolve ``1T w/ zgauss'' model. The temperature and normalization of the model were left free.
In the energy range of 5.4--9.0 keV, the energy scale uncertainty is $\pm0.3$ eV \citep{Eckart25}. 
Allowing the redshift of each line to vary within this $\pm 0.3$ eV range does not change the best-fit line widths or fluxes.
Since the Lyman-series doublets, with the exception of Fe \emissiontype{XXVI} Ly$\alpha$, were not resolved spectrally, the line widths and flux ratios were tied to the values of the corresponding Ly$\alpha_1$, as shown in table \ref{tab:line_list}.
The unresolved substructures of S \emissiontype{XV} He$\beta$, Ca \emissiontype{XIX} He$\beta$, Fe \emissiontype{XXV} He$\gamma$, and Fe \emissiontype{XXV} He$\delta$ were represented by a single $zgauss$ model.
The resultant normalization and line width of each line obtained by fitting are shown in table \ref{tab:line_list}.

Figure \ref{fig:line_ratio} shows the temperature dependence of line flux ratios in a CIE plasma (the $apec$ model). 
The excitation temperatures obtained from the line ratios of different transitions of the same ions 
(same in both the element and ionization state) 
are shown in the upper part of figure \ref{fig:line_ratio}. The ionization temperatures, which are the line ratios of different ionization stages of the same elements, are shown in the lower part of figure \ref{fig:line_ratio}. 
The derived excitation temperatures from the observed line ratios, such as S He$\beta$/$w$, Fe He$\beta_1$/$z$, and Fe He$\gamma$/$z$, are broadly consistent with the temperature derived from the Resolve ``1T w/ zgauss'' model. 
On the other hand, all the measured ratios give tighter constraints on the ionization temperatures.  

Figure \ref{fig:kT_line_ratio} summarizes the derived excitation and ionization temperatures from the observed line ratios. Under a CIE plasma, the excitation and ionization temperatures are expected to be the same. 
Although the excitation temperatures are consistent with the temperature derived from the Resolve ``1T w/ zgauss'' model, the ionization temperatures show a systematic increase with atomic numbers. A similar trend has been reported for the Perseus cluster observation with Hitomi \citep{2018PASJ...70...11H}, and these indicate the existence of the multi-phase gas in the cluster core. 

Some of the galaxy clusters observed with XRISM/Resolve observations exhibit Fe \emissiontype{XXVI} Ly$\alpha_2$/Ly$\alpha_1$ flux ratios that exceed the value predicted by a CIE plasma model (A2029: \cite{2025ApJ...982L...5X}, Coma: \cite{2025ApJ...985L..20X}). 
In the Centaurus cluster, the measured Fe \emissiontype{XXVI} Ly$\alpha_2$/Ly$\alpha_1$ flux ratio is consistent with the expected number for a CIE plasma within the statistical uncertainty. Whether using only the Ly$\alpha_1$ or the combined Fe \emissiontype{XXVI} Ly$\alpha_1$ $+$ Ly$\alpha_2$ flux, the ionization temperatures inferred from the line flux ratios are different from the resultant temperature derived from the Resolve ``1T w/ zgauss'' model, although only the Fe \emissiontype{XXVI} Ly$\alpha_1$ 
is adopted in the analysis of the line ratios in this paper. 

Figure \ref{fig:kT_Fe_wtoz} shows the measured line ratio of Fe \emissiontype{XXV} He$\alpha$ resonance ($w$) and forbidden ($z$). The observed $w/z$ ratio is smaller than that expected from the isothermal CIE plasma model, suggesting that the $w$ line is suppressed by resonant scattering, as reported in the central region of the Perseus cluster \citep{2018PASJ...70...10H}.

\subsection{Ion temperature measurements}
\label{sec:Tion}
As described in sections \ref{subsection:Thermal models}, \ref{sec:Line ratio diagnostics}, the CIE assumption seems to hold in the ICM of the Centaurus cluster core. Here, we further test this assumption by measuring the ion temperature and comparing it to the electron temperature.
Thanks to the high-resolution spectroscopy of Resolve, it is possible to directly and quantitatively estimate each component of the line broadening: the sum of a natural width, the instrumental resolution of the detector, the turbulent gas motion, and the thermal motion of ions. Since the natural width is much smaller than the other components, 
the line broadening is denoted as $\sigma_{v+\textup{th}}=\sqrt{\sigma_{v}^2 + \sigma_{\textup{th}}^2}$, where $\sigma_{v}$ is broadening due to turbulent gas motion and $\sigma_{\textup{th}}$ is broadening due to thermal motion of ions.  $\sigma_{\textup{th}}$ is expressed as 
$\sigma_{\textup{th}}=\sqrt{kT_{\textup{ion}}/m_{\textup{ion}}}$, where $T_{\textup{ion}}$ is the ion temperature, and $m_{\textup{ion}}$ is the ion mass. 
The line broadening due to thermal motion of ions depends on the ion mass, while the broadening due to turbulent gas motion is independent of the ion mass.

\begin{figure*}[t]
\begin{minipage}{0.5\textwidth}
 \begin{center}
 \includegraphics[width=80mm]{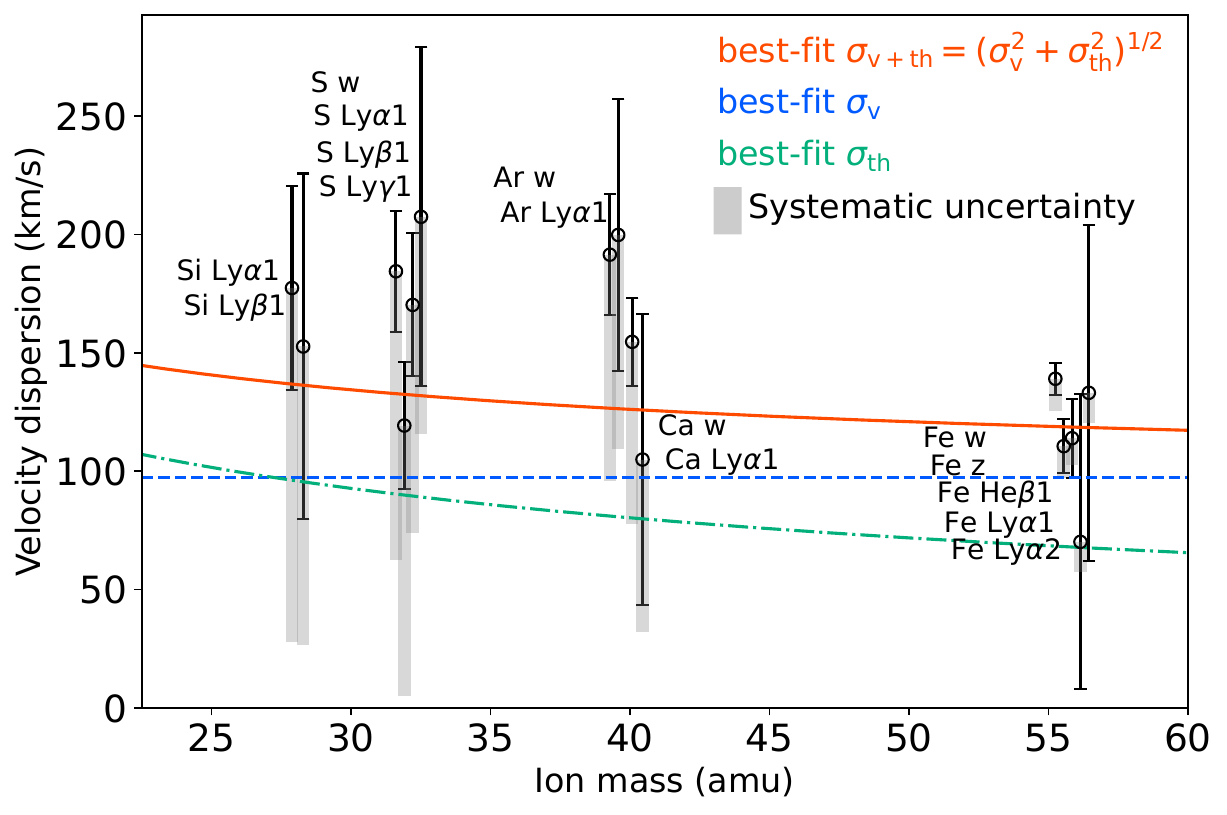}
 \end{center}
 \end{minipage}
 \begin{minipage}{0.5\textwidth}
 \begin{center}
 \includegraphics[width=70mm]{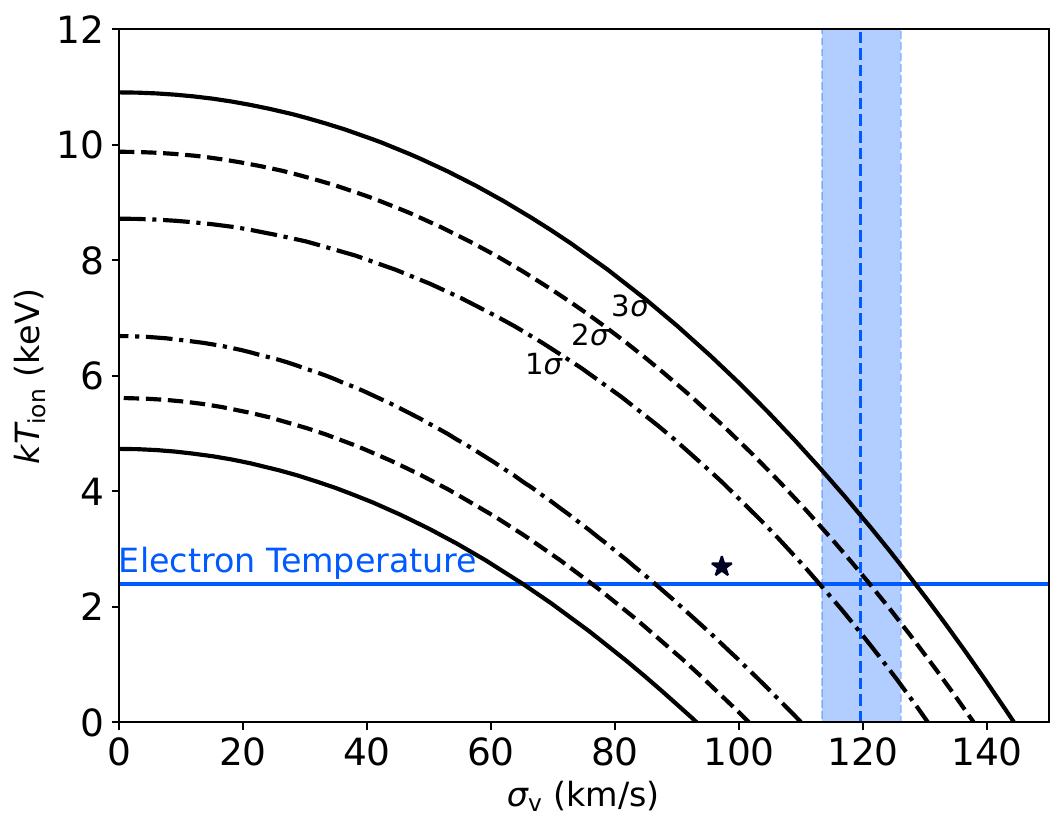}
 \end{center}
 \end{minipage}
\caption{ {\it Left:} The velocity dispersion of each line plotted against ion mass. The error bars for each line represent only statistical uncertainties. The gray-hatched boxes for each point correspond to systematic uncertainties of 0.3 eV for Fe and 1 eV for other elements. The fit was performed using uncertainties calculated by adding the statistical and systematic errors in quadrature. Data points from the same element are horizontally shifted for visibility. The red solid, blue dashed, and green dash-dotted lines represent the best-fitted $\sigma_{v+{\textup{th}}}$, $\sigma_{v}$, and $\sigma_{\textup{th}}$ values, respectively. {\it Right:} The confidence contour map of $\sigma_{v}$ and $kT_{\textup{ion}}$. The black dash-dotted, dashed, and solid curves indicate the 1$\sigma$, 2$\sigma$, and 3$\sigma$ confidence levels. The star-shaped marker shows the resultant best-fit value. The blue solid horizontal and blue dashed vertical lines represent the electron temperature and $\sigma_{v}$ values, respectively, derived from the ``1T w/ zgauss'' model.  {Alt text: Two-line graph. In the left panel, the y-axis shows the velocity dispersion, and the x-axis shows the ion mass. In the right panel, the y-axis shows $kT_{\textup{ion}}$ and the x-axis shows $\sigma_v$.}} 
\label{fig:kTion}
\end{figure*}

We estimated the velocities using the emission lines listed in table \ref{tab:line_list}, but the following lines were excluded from the analysis: (i) lines represent by a single $zgauss$ model for two lines,  (ii) lines for which only an upper limit on the line width was obtained, (iii) Ly$\alpha_2$, Ly$\beta_2$, He$\beta_2$ lines except for Fe \emissiontype{XXVI} Ly$\alpha_2$. 
The Si $w$ and $z$ lines were excluded because the statistics were inadequate to use them. As shown in the left panel of figure \ref{fig:kTion}, the observed line widths can be decomposed into the turbulent broadening (velocity dispersion) and thermal broadening for each ion at a certain temperature. As described in subsection \ref{subsection:Thermal models}, the velocity dispersions were derived from spectral fits using a thermal plasma model that includes the effect of the thermal broadening. Now we express the sum of the turbulent and thermal broadenings as a single velocity dispersion. Although the uncertainty in the full-width half maximum (FWHM) of the instrumental line-spread function (LSF, \cite{Leutenegger25}) results in errors in the velocity dispersion ($\pm 0.15$ eV at 6.7 keV and $\pm 0.06$ eV at 2.5 keV adds $\pm 2$ km s$^{-1}$ and $\pm 6$ km s$^{-1}$, respectively, to $\sigma_v = 120$ km s$^{-1}$) smaller than our statistical error of $\pm 7$ km s$^{-1}$, they are dominated by uncertainty in line broadening due to potential energy-scale misalignment across the pixels.  Aside from at 5.9 keV (where the energy scales are contrived to agree) and 6.5 keV (where the dispersion in energy-scale errors across pixels was determined to be 0.2 eV, see \cite{Simionescu26}), there is no available estimate for this potential broadening. Therefore, we have used the recommended values for the energy-scale uncertainty ($\pm 1$ eV below 5.4 keV and $\pm 0.3$ eV over the range 5.4--8.0 keV, \cite{Eckart25}) as a conservative estimate of possible broadening due to energy scale misalignment across the array.  Since line broadening results in determination of larger velocity dispersion, the associated velocity errors only extend to values below the best-fit value. The quadrature sum of statistical and systematic errors was taken into account in the fit shown in figure \ref{fig:kTion}, which shows the total velocity dispersion of the sum of the turbulence ($\sigma_v$) and thermal broadening ($\sigma_{\textup{th}}$). 
Since $\sigma_{v}$ was assumed to be constant for all lines, the contribution of broadening due to the thermal motion of ions is larger for elements with smaller atomic numbers relative to the observed line width. We examined the correlation of $\sigma_{v+\textup{th}}=\sqrt{\sigma_{v}^2 + \sigma_{\textup{th}}^2}$ to describe these line broadenings. Here, both $\sigma_{v}$ and $kT_{\textup{ion}}$ were freely varied.
The red solid, blue dashed, and green dash-dotted lines represent the resultant best-fit values of  $\sigma_{v+{\textup{th}}}$, $\sigma_{v}$, and $\sigma_{\textup{th}}$, respectively. 
The right panel of figure \ref{fig:kTion} shows the confidence contour map of $\sigma_{v}$ and $kT_{\textup{ion}}$. The derived ion temperature is consistent with the electron temperature from the Resolve ``1T w/ zgauss'' model.

\section{Discussion}
\subsection{Origins of deviation from the 1T CIE plasma} \label{subsection:Origins of deviation from the 1T CIE plasma}

\begin{figure}[t]
\begin{center}
\includegraphics[width=80mm]{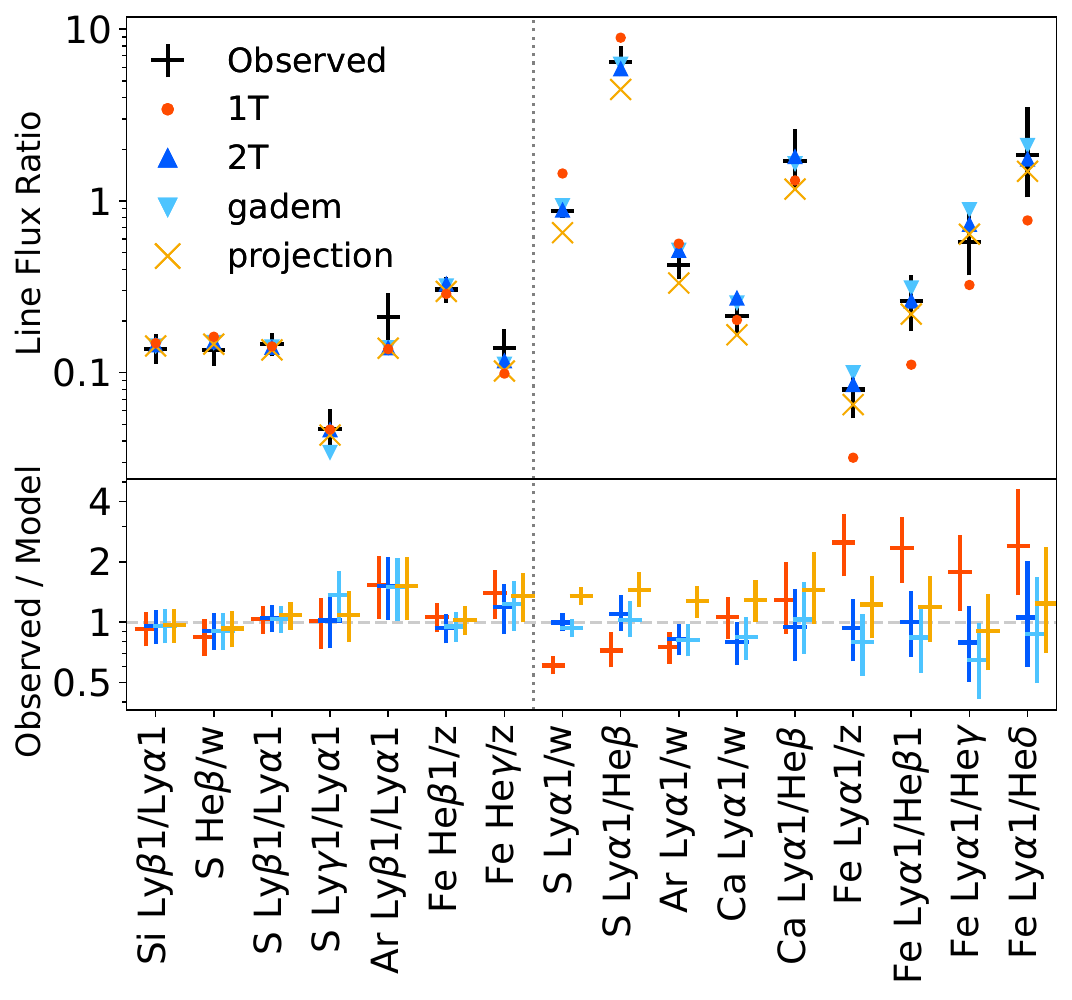}
\end{center}
\caption{Top panel shows comparisons between the observed line flux ratios and those predicted by the CIE model based on \atomdb{} 3.1.3. Black crosses represent the observed ratios. The red circle, blue triangle, light blue inverted triangle, and orange x-shaped markers correspond to the predictions from the 1T, 2T CIE plasma, $gadem$, and projection models, respectively, as described in the text. The bottom panel shows the ratio of observed to model-predicted line flux ratios. Vertical dotted line separates the ratios corresponding to the excitation temperatures and those corresponding to the ionization temperatures. {Alt text: One graph. The y-axis of the top panel shows the line flux ratio. The y-axis of the bottom panel shows the residuals between the data and the models. }} 
\label{fig:line_ratio_mo}
\end{figure}

As described in section \ref{subsection:Thermal models}, the observed Resolve spectrum is better represented by the two-temperature or, more generally, multi-temperature CIE plasma models than by the single-temperature model. The line ratio diagnostics also indicate the possible presence of multi-temperature plasma in the central region of the Centaurus cluster. Figure \ref{fig:line_ratio_mo} shows a comparison of the observed line flux ratios and the ratios predicted by the single-temperature, two-temperature thermal plasma, and $gadem$ models.
The line-flux ratios of each element predicted by the two-temperature plasma model show better agreement with the observations compared to those from the single-temperature model. 
The $gadem$ model also produces improved agreement in comparison to the single-temperature case. Some observed line ratios, such as Ar Ly$\beta_1$/Ly$\alpha_1$ and Ar Ly$\alpha_1/w$, indicated a tendency to exceed or fall below the predicted numbers of those models.
To quantify these differences, we calculated the $\chi^2$ values between the observed line flux ratios and those predicted by each model. The resultant $\chi^2$ values for the single-, two-, three-temperature CIE plasma, and $gadem$ models compared to the observed ratios are 46.2, 4.3, 4.0, and 6.4, respectively, for 16 degrees of freedom with no free parameters.
These $\chi^2$ values show that the two-temperature plasma model best reproduces the observed line-flux ratios. 

One plausible course for the deviation from a 1T plasma is the projection effect, which occurs when the radial temperature gradient from the center to the outskirts is integrated along the line of sight. To examine this possibility, we used the three-dimensional radial profiles of the temperature, density, and abundance derived from Chandra observation data with the SPEX $clus$ model (\cite{2025A&A...694A.149S}; see \cite{Anwesh 2025,Mernier2025submitted} for details of the analysis with the SPEX $clus$ model). For the outer region of the Centaurus cluster, we adopted the three-dimensional radial profiles from the Suzaku analysis of \citet{2013MNRAS.432..554W}. 
By integrating these three-dimensional radial profiles along the line of sight, we calculated the line flux ratios, which are shown as the orange x-shaped markers  (``projection'') in figure \ref{fig:line_ratio_mo}.  The projection model yields a $\chi^2$ value of 11.4. The resultant $\chi^2$ value is significantly smaller than that predicted by the 1T model. Although the $\chi^2$ value from the projection model is higher than those predicted by the 2T and gadem models, the observed line flux ratios, especially for Fe, are well represented by the projection model. On the other hand, a slight residual is observed in the S line ratio. This may be caused by uncertainties in the projection model based on the Fe reference from Chandra observations, and/or by the influence of the low-temperature component indicated by the RGS analysis as described in section \ref{subsec:rgs}. When the S Ly$\alpha_1$/$w$, Ly$\alpha_1$/He$\beta$ line ratios were excluded, the $\chi^2$ value for the projection model $\chi^2$ value was 4.7 for 14 degrees of freedom, consistent with those predicted by the 2T and gadem models. Therefore, we concluded that the projection effect adequately explains the observational results with XRISM/Resolve.

The projection effect also reasonably explains the tendency for elements with lower atomic numbers to have lower ionization temperatures. In addition, the line widths of emission lines from the lighter elements tend to be broader than those from the heavier ones as described in section \ref{sec:Tion}, since the emission-measure weighted average fluxes of the lines from the lighter elements should be biased to the cool gas in the line of sight, which is confirmed by the RGS observation. Therefore, the observed line ratios from the Resolve observations are consistent with the previous Chandra and XMM-Newton observations.

\subsection{Comparison with Previous Observations of Other Galaxy Clusters}

\citet{2018PASJ...70...11H} compared the observed and model-predicted Fe line ratios, which provide excitation temperatures, in the core of the Perseus cluster. In this study, we compared not only the Fe ratios but also the line ratios of Si, S, and Ar, which also provide excitation temperatures.  
As shown in figure \ref{fig:line_ratio_mo}, the observed Fe line ratios, which are used to derive the ionization temperature, tend to be higher than those predicted by the single-temperature CIE model. This trend is consistent with that observed in the Perseus cluster \citep{2018PASJ...70...11H}. \citet{2018PASJ...70...11H} reported that the observed Fe Ly$\alpha$/$z$, Fe Ly$\alpha$/He$\beta$, and Fe Ly$\alpha$/He$\gamma$ ratios were higher than those predicted by the two-temperature CIE plasma model. In contrast, in this work, the observed Fe Ly$\alpha_1$/$z$, Fe Ly$\alpha_1$/He$\beta_1$, and Fe Ly$\alpha_1$/He$\gamma$ ratios show good agreement with the 2T CIE model predictions.

\citet{2025PASJ...77S.254S} performed temperature measurements for Abell~2029 using line diagnostics based on various Fe lines. Since the ICM in Abell~2029 is hotter than that in the Centaurus cluster, the Fe emission lines are stronger. In the central and northern regions, they resolved the cooler gas component lower than the temperature derived from the single-temperature CIE model. They interpreted this as evidence for the presence of multi-phase gas. In the pointing located further to the north, The temperatures from the line ratios were consistent with the values predicted by the single-temperature CIE model. For the Centaurus cluster as well, it is possible that in the outer regions the temperatures derived from line diagnostics agree with those predicted by the single-temperature CIE model.

For the Perseus cluster observation \citep{2018PASJ...70...11H}, both $\sigma_{v}$ and $kT_{\textup{ion}}$ were set as free parameters to derive the ion temperature. On the other hand, for the Centaurus cluster, when both $\sigma_{v}$ and $kT_{\textup{ion}}$ were allowed to vary freely, the best-fit value of $\sigma_{v}$ became significantly smaller than that obtained from the Resolve ``1T w/ zgauss'' model. Therefore, we fixed $\sigma_{v}$ and allowed only $kT_{\textup{ion}}$ to vary (section~\ref{sec:Tion}). 
Due to its lower temperature compared with the Perseus cluster, we expect that the uncertainty in the velocity dispersion, $\sigma_{v}$, in the Centaurus cluster is large. Since thermal broadening contributes only a small fraction to the total line width, the thermal and turbulent components cannot be separated.

\subsection{Resonant Scattering}

\begin{figure}[t]
\begin{center}
\includegraphics[width=80mm]{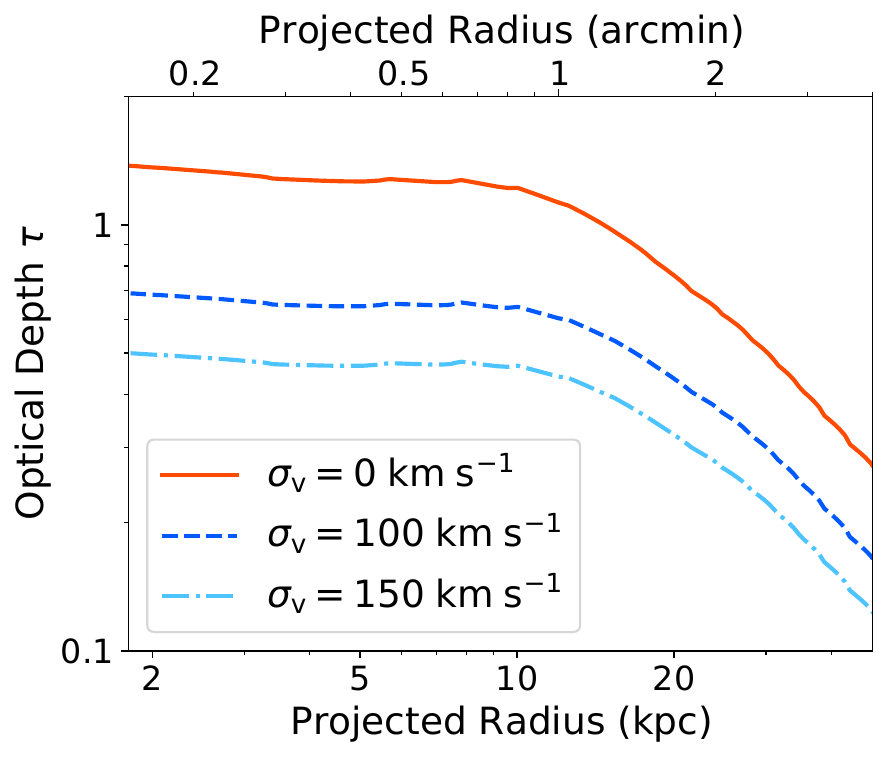}
\end{center}
\caption{The radial profile of optical depth of the Centaurus cluster. The solid lines in red, blue, and cyan indicate the results for velocity dispersion of 0\,km\,s$^{-1}$, 100\,km\,s$^{-1}$, and 150\,km\,s$^{-1}$, respectively. {Alt text: One line graph. The y-axis shows the optical depth. The x-axis shows projected radius. }} 
\label{fig:OpticalDepth}
\end{figure}

The effect of resonant scattering is most prominently seen in the Fe \emissiontype{XXV} He$\alpha$ resonance line, which is the strongest line in the Resolve spectrum. Following the procedure of \citet{2018PASJ...70...10H}, we calculated the optical depth based 
on the Chandra data analysis with the SPEX $clus$ model \citep{Anwesh 2025}, 
and the results are shown in figure \ref{fig:OpticalDepth}. Under the assumption of a velocity dispersion of 100\,km\,s$^{-1}$, the optical depth of the Fe \emissiontype{XXV} He$\alpha$ resonance line reaches $\sim 0.7$ within 1 arcmin from the cluster center.
Therefore, resonant scattering can be diagnosed by deviations of the observed Fe \emissiontype{XXV} $w/z$ ratio from the CIE prediction. 

As demonstrated in section \ref{sec:Line ratio diagnostics}, the observed $w/z$ ratio of $2.1^{+0.3}_{-0.2}$ is lower than the values expected by any of the isothermal CIE models used for comparison.
At the temperature of $2.39_{-0.02}^{+0.02}$ keV derived from the Resolve ``1T w/ zgauss'' model, the predicted intrinsic $w/z$ ratio is about $2.6$. 
The multi-temperature ``projection'' model also yields a similar value of $\sim$2.5.
If the deficit is interpreted as the result of resonant scattering, the Fe \emissiontype{XXV} He$\alpha$ $w$ line is suppressed by approximately $20\%$.
Although the difference between the observed and predicted $w/z$ ratios from the isothermal CIE prediction is within $2\sigma$-level significance, it provides a hint of resonant scattering in the central region of the Centaurus cluster.

Resonance scattering not only suppresses spectral lines at optical depths $\geq 1$, but also appears to enhance the wings of these spectral line profiles when attempting to describe them using a Gaussian model under conditions of limited energy resolution and insufficient statistics \citep{2018PASJ...70...10H}.
The observed line width was larger for Fe \emissiontype{XXV} He$\alpha$ $w$ than $z$, where the difference was about $2\sigma$ (table \ref{tab:line_list}).
As noted in section \ref{sec:Tion}, the observed line width of each emission line involves systematic uncertainties due to the uncertainty in the FWHM of the instrumental line-spread function. With this systematic uncertainty, the observed width of the Fe \emissiontype{XXV} He$\alpha$ $w$ line remains larger than the width of the $z$ line. Although other possibilities, such as a charge exchange effect, recombining plasma, or the uncertainties in the atomic excitations in \citet{2018PASJ...70...10H} could not be entirely ruled out, the resonant scattering provides the most plausible explanation for these observational results, namely the $w/z$ line ratio and the broader line width of the $w$ line. 
The observed Fe $w$ line suppression is consistent with the turbulent motions inferred from the line broadening, and hence the motions are consistent with being isotropic. This indicates that the line broadening is driven by small scale motions rather than a superposition of bulk flow in the line of sight.
Consequently, we found the indication of resonant scattering in the Centaurus cluster core; the significance was approximately at the 2$\sigma$ level.

\section{Summary and Conclusion}

We found the temperature structure of the Centaurus cluster core with XRISM/Resolve. In the spectral analysis in the 1.8--12 keV energy band, the two-temperature and DEM models better represented the observed spectrum of the cluster core than the isothermal model. 
Even if the cool component ($kT <$1 keV) derived from the XMM-Newton/RGS analysis is added in the spectral model, the resultant parameters of the Resolve spectral analysis are not affected. We performed line diagnostics using the line flux ratios of Si, S, Ar, Ca, and Fe to determine the excitation and ionization temperatures. The excitation temperatures derived from the line ratios other than Ar Ly$\beta_1$/Ly$\alpha_1$ agree with the derived temperature from the Resolve ``1T w/ zgauss'' model. On the other hand, the resultant ionization temperatures revealed the deviation from the 1T CIE plasma model. The derived ionization temperature from the line flux ratios exhibited a tendency to increase in proportion to the atomic mass. A similar trend was reported in the central region of the Perseus cluster observed with Hitomi/SXS. As a result, the derived temperature from the Resolve observation above 2 keV indicates the multi-temperature gas, and the temperature structure of the Centaurus cluster core can be simply explained by the projection effect.
The observed Fe \emissiontype{XXV} $w$/$z$ ratio was lower than the value predicted by a CIE plasma model. Compared with the isothermal CIE plasma model, the Fe $w$ line was suppressed by approximately 20\%, and the line width was broader than that of the Fe $z$ line. These results suggest the presence of resonance scattering in the core of the Centaurus cluster, with a significance of approximately 2$\sigma$ level.

\begin{ack}
We would like to thank Drs. Makoto S. Tashiro, Yukikatsu Terada, and Satoru Katsuda for constructive comments. This work was supported by JSPS Core-to-Core Program, (grant number:JPJSCCA20220002). NW and TP were supported by the GACR EXPRO grant No. 21-13491X. The authors are also grateful to R. Mushotzky and C. A. Kilbourne for valuable comments and discussions.
\end{ack}

\appendix

\section{Details of the Resolve spectral analysis}
\label{app:spectral_analysis}
\begin{figure*}[t]
\begin{minipage}{0.5\textwidth}
 \begin{center}
 \includegraphics[width=80mm]{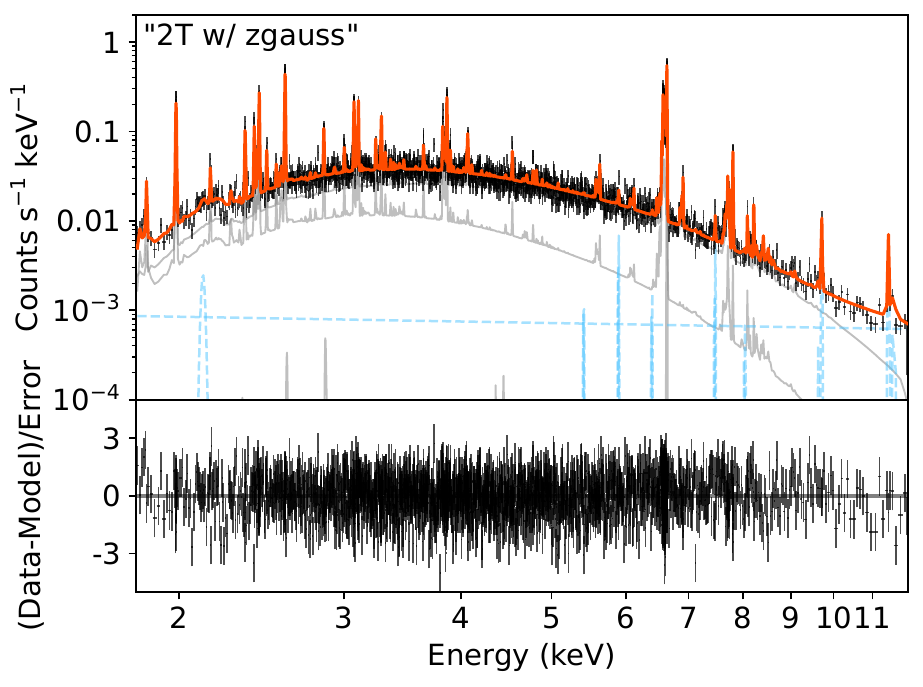}
 \end{center}
 \end{minipage}
 \begin{minipage}{0.5\textwidth}
\begin{center}
\includegraphics[width=80mm]{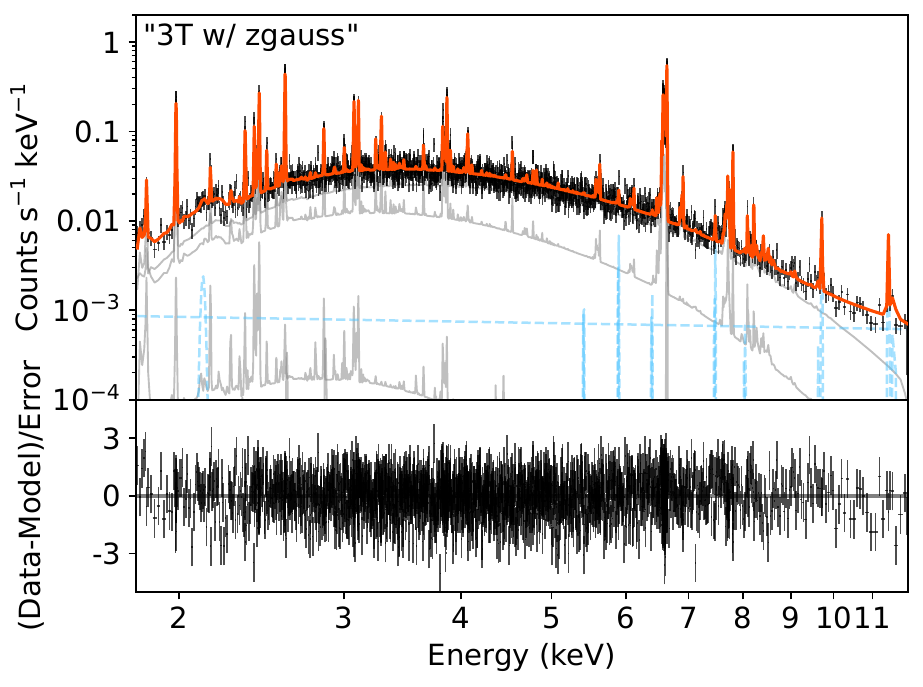}
\end{center}
\end{minipage}
 \caption{Spectral fits with the Resolve ``2T w/ zgauss'' (left) and the ``3T w/ zgauss'' (right) models in the 1.8--12 keV band. The red solid line corresponds to the best-fit parameter of each model. The light gray lines indicate each temperature and Gaussian component for the best-fit models. The light blue dashed line indicates the NXB model. The lower panels show the residuals between the data and the model. {Alt text: Two line graphs. In the top panels, the y-axis shows the count per second per kilo-electron volt. In the bottom panel, the y-axis shows the residual between the data and the model.}
 }\label{fig:appendix_spec4}
\end{figure*}

\begin{figure*}[t]
\begin{minipage}{1.0\hsize}
\begin{center}
\includegraphics[width=145mm]{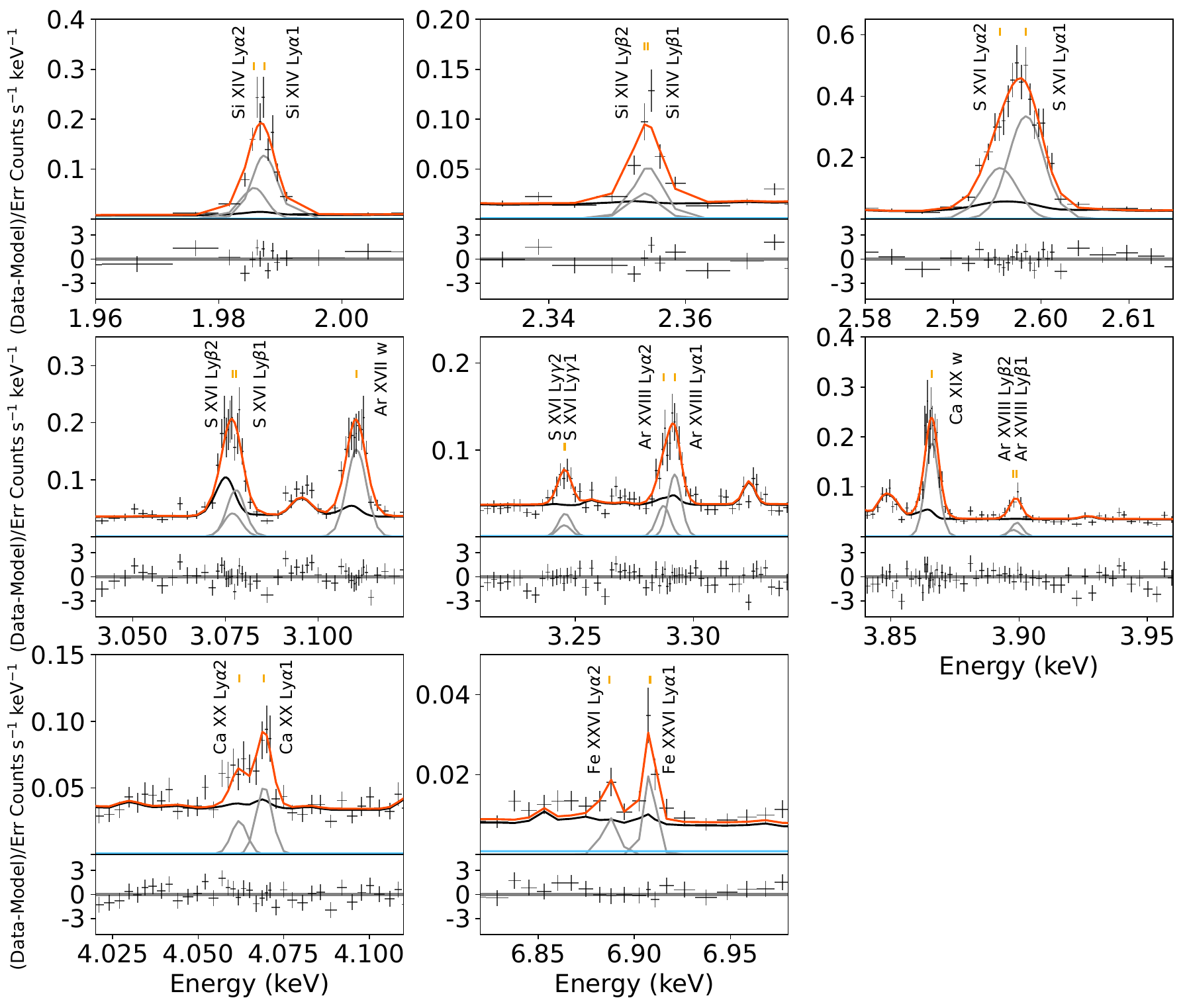}
\end{center}
\end{minipage}
\caption{Resultant spectral fits and their residuals around H-like emission lines from each element. The spectra were grouped for display purposes. The red solid and gray lines show the best-fit model and each $zgauss$ component, respectively. The light blue lines indicate the NXB model. The lower panel shows the residuals between the data and the best-fit model. {Alt text: Eight line graphs. In the upper panel of each graph, the y-axis shows the count rate per second per kilo–electron volt. In the lower panel of each graph, the y-axis shows the residuals between the data and the model. The x-axis in all panels shows the energy.}} 
\label{fig:appendix_spec1}
\end{figure*}

\begin{figure*}[t]
\begin{minipage}{1.0\hsize}
\begin{center}
\includegraphics[width=145mm]{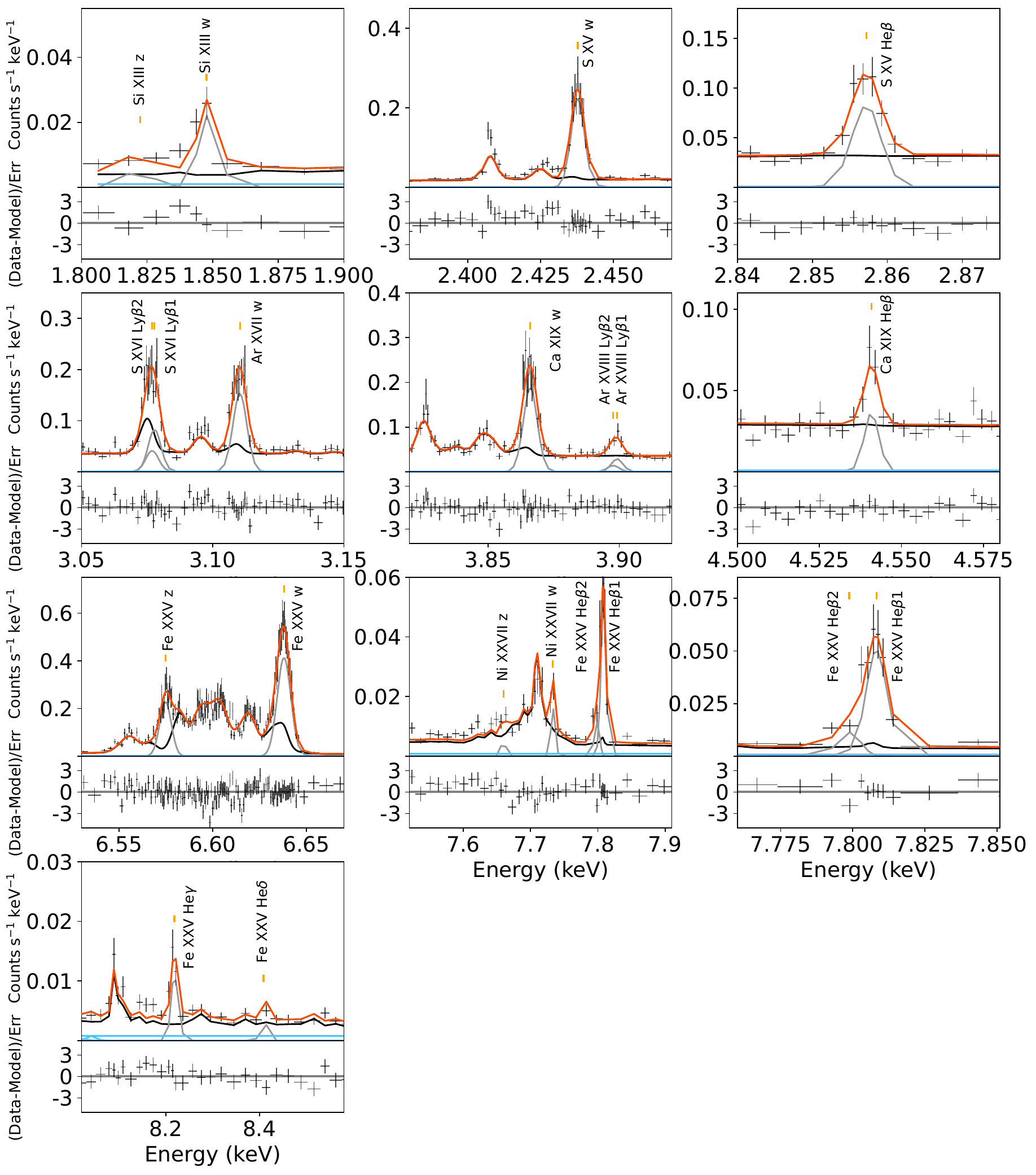}
\end{center}
\end{minipage}
\caption{Same as figure \ref{fig:appendix_spec1}, but for He-like emission lines from each element. {Alt text: Eight line graphs. In the upper panel of each graph, the y-axis shows the count rate per second per kilo–electron volt. In the lower panel of each graph, the y-axis shows the residuals between the data and the model. The x-axis in all panels shows the energy.}} 
\label{fig:appendix_spec2}
\end{figure*}

Figure \ref{fig:appendix_spec4} shows the wide-band spectral fits and their residuals with the Resolve ``2T w/ zgauss'' and ``3T w/ zgauss'' models in the 1.8--12 keV band. The lower temperature component below 1 keV derived from the XMM-Newton/RGS analysis was negligible for the Resovle spectral fit analysis above 2 keV. Figures \ref{fig:appendix_spec1} and \ref{fig:appendix_spec2} show the resultant fits and their residuals around H- and He-like emission lines from each element. Resultant parameters for each line are shown in table \ref{tab:line_list}.

\section{Comparison of the projected line ratios between \atomdb{} and the SPEX}
\begin{figure}[t]
\begin{center}
\includegraphics[width=80mm]{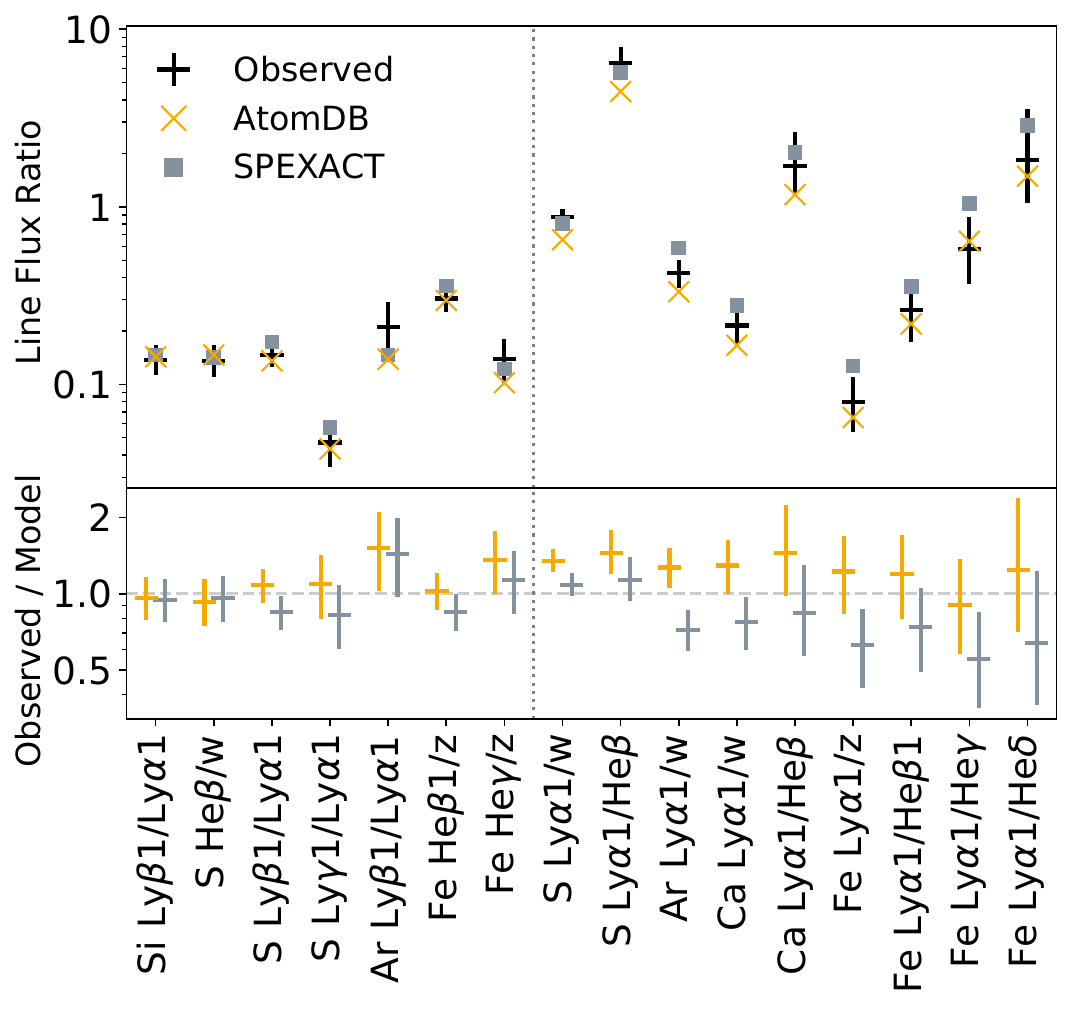}
\end{center}
\caption{Comparisons between the observed line flux ratios and the predicted ratios by the \atomdb{} and SPEX. Black crosses show the observed line ratios. The orange x-shaped and gray square markers correspond to the predictions from the projection model based on the \atomdb{} 3.1.3 and SPEXACT 3.08.01. The bottom panel shows the ratio of observed to model-predicted line ratios. Vertical dotted line separates the ratios corresponding to the excitation temperatures and those corresponding to the ionization temperatures. {Alt text: One graph. The y-axis of the top panel shows the line flux ratio. The y-axis of the bottom panel shows the residuals between the data and the models. }} 
\label{fig:line_ratio_mo_pro}
\end{figure}

The projected line ratio models derived from the $clus$ model based on the Chandra and Suzaku observations represent the observed line ratios as described in section \ref{subsection:Origins of deviation from the 1T CIE plasma}. We compared the line ratios derived from the \atomdb{} and the SPEX, although the $clus$ model was developed based on SPEX. Figure \ref{fig:line_ratio_mo_pro} shows a comparison of the predicted line ratios derived from the \atomdb{} and SPEX, and the observed line ratio is the same as those in figure \ref{fig:line_ratio_mo}. These two profiles are generally consistent with each other, although slight differences are seen between \atomdb{} and SPEXACT in the line flux ratios of the S and Fe lines.

\end{document}